\documentclass{emulateapj}

\usepackage{natbib}
\pdfoutput=1
\usepackage[monochrome]{color}
\usepackage{rotating}
\usepackage{mathtools}

\shorttitle{ECO and RESOLVE: Disk Growth and Environment}
\shortauthors{Moffett et al.}

\begin{document}


\title{ECO and RESOLVE: Galaxy Disk Growth in Environmental Context}


\author{Amanda J. Moffett \altaffilmark{1,2}, Sheila J. Kannappan \altaffilmark{1}, Andreas A. Berlind \altaffilmark{3}, Kathleen D. Eckert \altaffilmark{1}, David V. Stark \altaffilmark{1}, David Hendel \altaffilmark{4}, Mark A. Norris \altaffilmark{5}, and Norman A. Grogin \altaffilmark{6}}

\altaffiltext{1}{Dept. of Physics \& Astronomy, University of North Carolina, Phillips Hall, CB 3255, Chapel Hill, NC 27599}

\altaffiltext{2}{International Centre for Radio Astronomy Research (ICRAR), University of Western Australia, M468, 35 Stirling Highway, Crawley, WA 6009, Australia}

\altaffiltext{3}{Dept. of Physics \& Astronomy, Vanderbilt University, 6301 Stevenson Center, Nashville, TN 37235}

\altaffiltext{4}{Dept. of Astronomy, Columbia University, Mail Code 5246, 550 West 120th Street, New York, NY 10027}

\altaffiltext{5}{Max-Planck-Institut fur Astronomie, Konigstuhl 17, D-69117, Heidelberg, Germany}

\altaffiltext{6}{Space Telescope Science Institute, 3700 San Martin Drive, Baltimore, MD 21218}


\begin{abstract}
We study the relationships between {\color{red}{galaxy environments}} and galaxy properties related to disk (re)growth, considering two highly complete samples
that are approximately baryonic mass limited into the high-mass dwarf
galaxy regime, the Environmental COntext (ECO) catalog (data release
herein) and the B-semester region of the REsolved Spectroscopy Of a
Local VolumE (RESOLVE) survey. We quantify galaxy environments using both group identification and smoothed galaxy density field methods. We use by-eye and quantitative morphological classifications plus atomic gas content measurements and estimates. We find that blue early-type (E/S0) galaxies, gas-dominated galaxies, and UV-bright disk host galaxies all become distinctly more common below group halo mass $\sim10^{11.5} M_{\odot}$, implying that this low group halo mass regime may be a preferred regime for significant disk growth activity. We also find that blue early-type and blue late-type galaxies inhabit environments of similar group halo mass at {\color{red}{fixed}} baryonic mass, consistent with a scenario in which blue early types can regrow late-type disks. {\color{red}{In fact, we find that the only significant difference in the typical group halo mass inhabited by different galaxy classes is for satellite galaxies with different colors, where at fixed baryonic mass red early and late types have higher typical group halo masses than blue early and late types. More generally, we argue that the traditional morphology-environment relation (i.e., that denser environments tend to have more early types) can be largely attributed to the morphology-galaxy mass relation for centrals and the color-environment relation for satellites.}}

\end{abstract}





\section{Introduction}

  For decades, astronomers have observed that the properties of galaxies in the local universe, including appearance, star formation rate, and gas content, depend on the surrounding environment (e.g., as reviewed by \citealp{BB04}; \citealp{BG06}). Galaxies of different morphological types, in particular, have long been seen to preferentially congregate in different environments (e.g., \citealp{HH31}; \citealp{DG76}). \citet{Dressler80} reported the so-called ``morphology-density relation,'' whereby E/S0 fractions increase with increasing environmental density within rich clusters while spiral fractions decrease. \citet{PG84} showed that a similar relationship between galaxy morphology and local density also holds in the lower-density group environment. However, several authors found the conflicting result that significant variation in morphology with environment exists only in the richest clusters (e.g., \citealp{Maia90}; \citealp{Whitmore95}). Notwithstanding these early disagreements, the original \citet{PG84} observation of a morphology-density relation extending into less-rich environments has since been corroborated by a variety of other authors (e.g., \citealp{Tran01}; \citealp{HP03}; \citealp{Goto03}; \citealp{Calvi12}). 

Recently, it has also been noted that the mass ranges of galaxies considered can have a significant impact on the observed form of the morphology-density relation (e.g., \citealp{Bamford09}; \citealp{Calvi12}; \citealp{WE12}). While \citet{Drinkwater2001} observe a traditional morphology-density relation in the dwarf galaxy population within a single cluster, other studies find that the morphology-density relation takes on different forms for low-to-intermediate mass galaxies. \citet[hereafter KGB]{KGB} find suggestive evidence that low-to-intermediate stellar mass E/S0s occupy low density environments similar to those of spirals at the same masses, while \citet{Calvi12} find that morphologies for intermediate mass galaxies are not closely related to environment, except in clusters. Interestingly, \citet{Hogg03} find that the mean environmental overdensity for red, typically assumed to be early-type (E/S0), galaxies reaches a minimum at intermediate masses/luminosities (around $L_{*}$). The conflation of morphology and color, of course, complicates interpretation of the Hogg et al. result, since both optically red, ``passive'' spiral galaxies (e.g., \citealp{vdbergh76}; \citealp{Couch98}; \citealp{Dressler99}; \citealp{Poggianti99}) and optically blue E/S0 galaxies (e.g., KGB; \citealp{Schawinski09}) are known components of the galaxy population.

\begin{figure*}
\epsscale{1.12}
\plotone{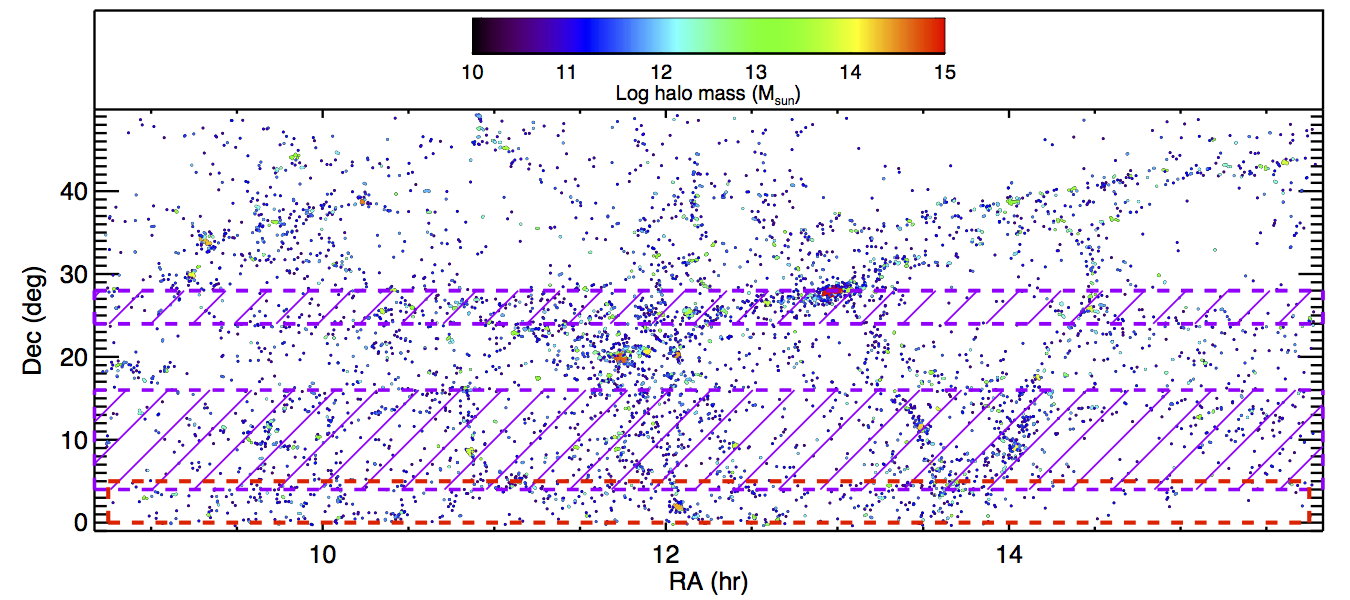}
\caption{The ECO catalog region in sky coordinates, with {\color{red}{the ECO mass-limited sample represented by points color-coded}} according to group halo masses estimated as described in \S \ref{envmet}. The RESOLVE-A region is outlined in red, and the region of overlap with the ALFALFA $\alpha.40$ catalog \citep{A40cat} is indicated by the purple crosshatched strips.}
\label{skycov}
\end{figure*}

It is likely that the existence of such non-traditional color-morphology pairings in the galaxy population is an important factor driving the observation that trends in galaxy color (or more direct star formation property measures) versus environment can differ from observed morphology-density trends. Beginning with the analysis of \citet{Kennicutt83}, it has often been observed that cluster galaxies show typically lower levels of star formation than galaxies in less rich environments. Several authors have subsequently found that such star formation or color trends with environment cannot be completely explained by the presence of morphology-density trends (e.g., \citealp{KK98}; \citealp{Lewis02}; \citealp{CZ05}; \citealp{Welikala09}). Others have taken this conclusion further, inferring that trends in star formation/color are actually more closely linked to environmental conditions than are trends in structural/morphological parameters (e.g., \citealp{Blanton03}; \citealp{Kauffmann04}; \citealp{Blanton05}; \citealp{Quintero06}; \citealp{Bamford09}; \citealp{Skibba09}).

Galaxy gas content is another property that has been observed to share a close link to both star formation and the ambient environmental conditions around galaxies. First observed by \citet{DL73} for galaxies in the Virgo Cluster, the result is now well established that cluster galaxy populations typically display lower levels of atomic gas than do similar populations in less dense environments (e.g., \citealp{GH83}; \citealp{Haynes84}; \citealp{GH85}; \citealp{Gavazzi87}; \citealp{Bravo00}; \citealp{Solanes01}; see also review by \citealp{vanGorkom04}). Likewise, cluster galaxies are typically observed to have HI gas disks that are less extended than those in lower density environments (e.g., \citealp{GH83}; \citealp{Warmels88}; \citealp{Cayatte94}; \citealp{Bravo00}). Galaxy gas properties can apparently be affected by the conditions present in lower density environments as well, e.g., as shown in the simulations of \citet{KM08} where ``strangulation'' or stripping of a hot galaxy halo gas component, which could otherwise cool to provide a cold gas reservoir, in a larger potential is effective in a group environment with halo mass $\sim10^{12.9}\, M_{\odot}$. Still further down the environmental density scale, gas-rich galaxies have been observed to be one of the most weakly clustered galaxy populations, that is, typically found in the lowest density environments (\citealp{Basilakos07}; \citealp{Meyer07}; \citealp{Martin12}; \citealp{Li12}). This observation could be related to the finding from multiple theoretical studies that gas accretion into galaxies, whether in a ``cold'' or ``hot'' mode, is most effective where group halo masses are low (e.g., \citealp{BD03}; \citealp{Keres05}; \citealp{DB06}; \citealp{Keres09}; \citealp{Nelson13}). 

In the prevailing hierarchical model of galaxy evolution, galaxies are thought to experience morphological transformations not just from late to early type, for example through merging/quenching processes, but also potentially from early to late type, through a disk \emph{regrowth} process that may be enabled by gas accretion (e.g., \citealp{Barnes02}; \citealp{SM02}; \citealp{Governato07}). Observationally, the operation of such a disk regrowth process is difficult to confirm. 

\begin{figure*}
\plotone{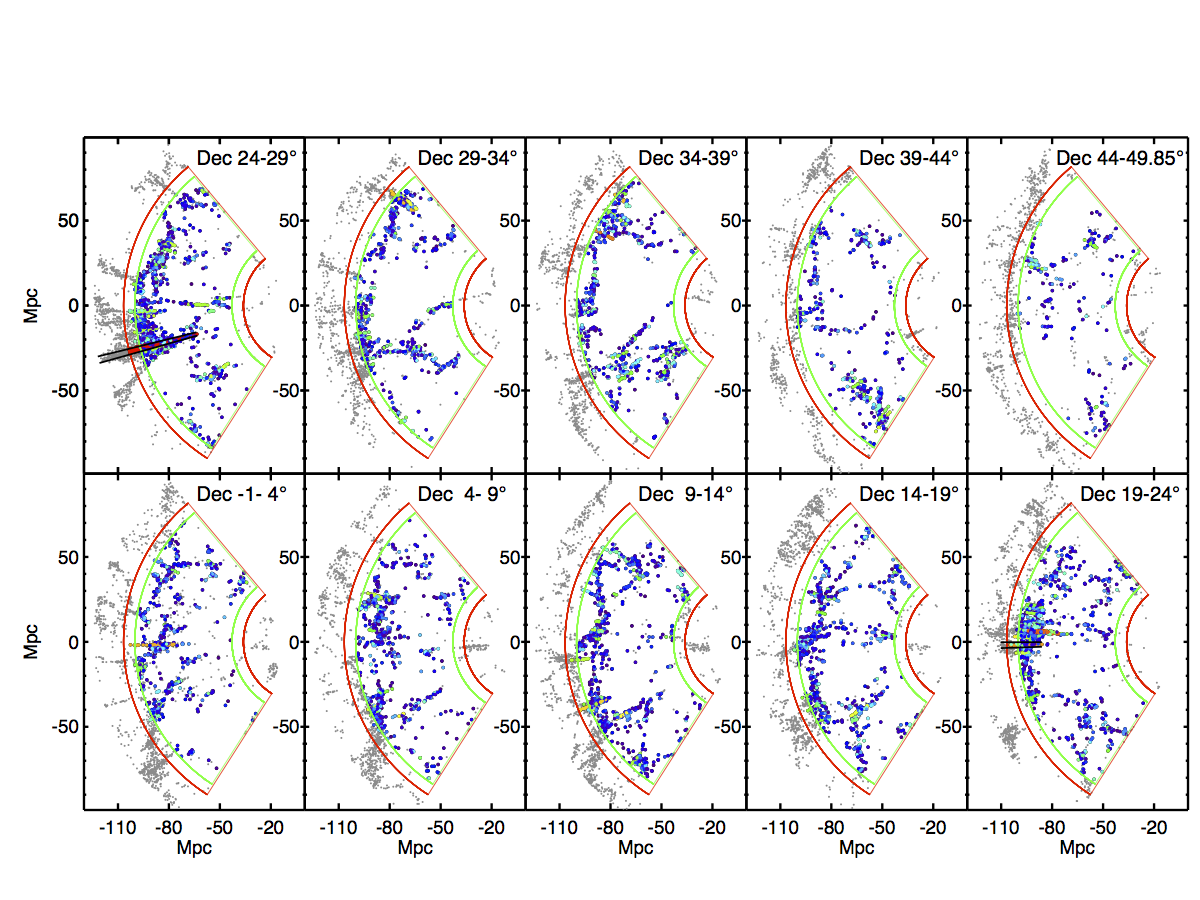}  
\caption{The ECO catalog region in RA vs.\ line-of-sight distance coordinates, in slices of $\sim$5 degrees in Dec, increasing from the bottom left panel. Galaxies included in the final ECO {\color{red}{mass-limited}} sample analyzed here are indicated by dots color-coded according to their group halo masses as in Fig.\ \ref{skycov}, while galaxies outside this sample but present in our merged parent redshift catalog are indicated by grey dots. The outer limits of the ECO catalog ``buffer'' region are outlined in red, and the region of ECO we consider interior to this buffer is indicated in green. {\color{red}{Black lines in the upper left and bottom right panels indicate the approximate line-of-sight extent of two groups that extend significantly outside the ECO region and are subject to boundary completeness correction factors as described in \S \ref{ccs}. }} }
\label{racz}
\end{figure*}

However, several hints have recently emerged that such a scenario is plausible. One such hint lies in the existence of the blue or ``blue-sequence'' E/S0 population, consisting of morphologically early type galaxies that lie on the blue sequence in color-stellar mass space. Blue-sequence E/S0s are typically found in non-cluster environments and exist primarily at stellar masses less than $\sim10^{10.5}\, M_{\odot}$ (the bimodality mass of \citealp{Kauffmann03}, hereafter $M_{b}$) but are most common below the ``gas-richness threshold'' stellar mass of $\sim10^{9.7} M_{\odot}$ (KGB). \citet[hereafter K13]{K13} argue that the bimodality and gas-richness threshold mass scales mark two distinct transition points between galaxy refueling regimes, with galaxies below the threshold scale typically experiencing high levels of external gas accretion and stellar mass growth. Consistent with this picture, low-mass blue E/S0s contain sufficient gas reservoirs and specific star formation rates to allow the growth of evolutionarily significant disk structures on relatively short timescales (KGB; \citealp{Wei10}). Another hint of disk regrowth in E/S0s is the observation of extended UV emission, associated with recent star formation, around a number of nearby galaxies with early-type morphology (e.g., \citealp{Donovan09}; \citealp{CH09}; \citealp{Thilker10}; \citealp{Me12}). \citet{Me12} observe that such extended UV structures are common in low-to-intermediate mass early-type galaxies and that a particular class of ``UV-Bright'' or UV-B disk galaxies is marked by a high potential for ongoing star formation. Linking these two populations together, UV-B disks are also strongly associated with the low-mass, blue-sequence E/S0 population, supporting the idea that these galaxies may be engaged in disk regrowth. \citet{Stark13} report evidence for gas as well as stellar disk regrowth in low-mass, post-starburst E/S0 galaxies.

In this contribution, we employ two unusually complete volume-limited galaxy samples, both of which extend into the ``high-mass'' dwarf galaxy regime (reaching baryonic masses $\sim10^{9.3} M_{\odot}$), to probe disk (re)growth in a variety of environments. We seek to answer three major questions. (1) Does environment play a role in enabling gas or stellar disk growth? (2) Does the morphology-density, or more generally morphology-environment, relation behave as might be expected if disk regrowth is effective in transforming galaxy morphology? In particular, are the typical environments for blue early- and late-type galaxies similar in the galaxy mass regimes in which disk regrowth occurs? (3) Are there specific group halo mass scales implicated in evidence for disk regrowth?

We address these questions in part by examining the detailed form of the morphology-environment relation, including possible variations with galaxy mass scale and central/satellite designations. In addition, we consider an alternative way of formulating a morphology-environment relation. If the traditional formulation of the morphology-environment relation can be considered to quantify the probability of a galaxy exhibiting a particular morphology given some environment, P(M$|$E), then an alternative way to frame this relation is to quantify the probability of a galaxy inhabiting a particular environment given its morphology, P(E$|$M). This alternative formulation provides a useful framework for understanding the typical environments of galaxies with different morphologies. We further examine the typical environments of different classes of galaxies linked to disk growth, including blue-sequence early types, galaxies with substantial atomic gas reservoirs, and early-type galaxies that display recent UV-detected disk star formation.

\begin{figure*}
\epsscale{1.1}
\plottwo{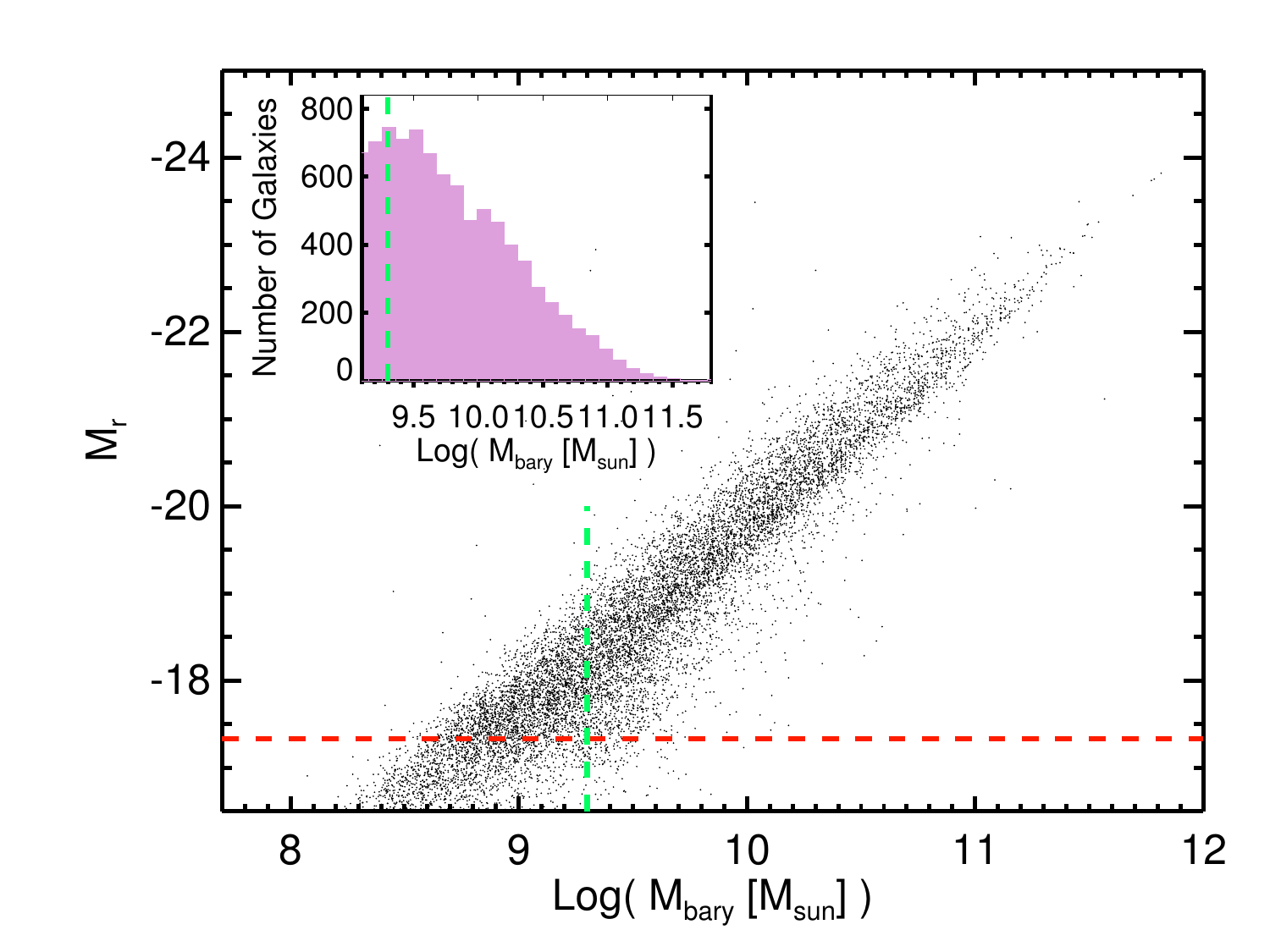}{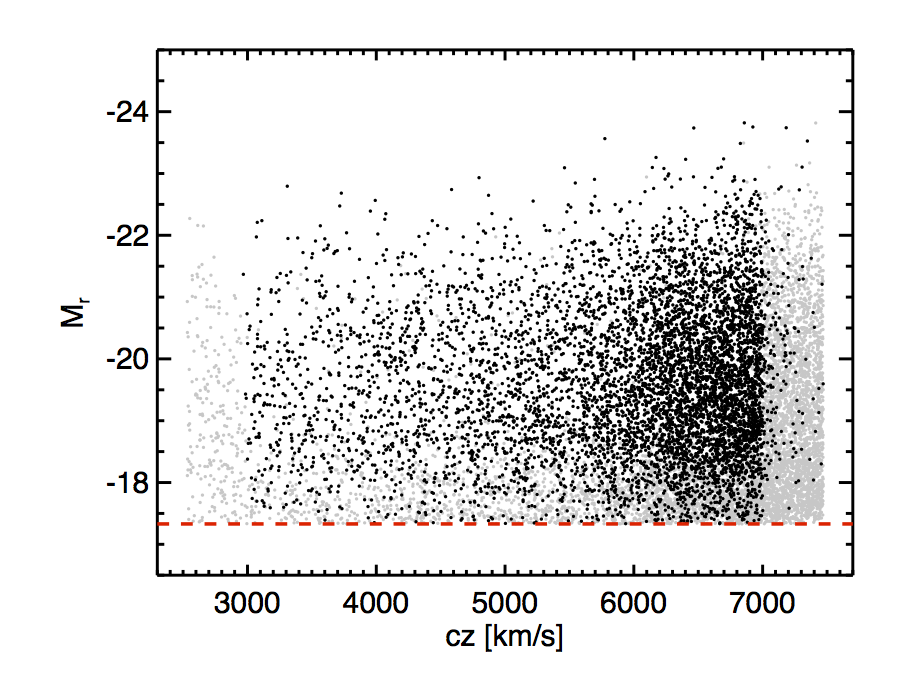}
\caption{Illustration of the selection of the ECO catalog.  Panel \emph{a} shows the $M_{r}$ and $M_{bary}$ distributions of the initial ECO catalog (dots and purple inset $M_{bary}$ histogram), where the horizontal red line indicates the $M_{r} < -17.33$ redshift completeness limit and the vertical green lines indicate the final mass cut we adopt to create an approximately baryonic mass limited sample to $M_{bary} > 10^{9.3} M_{\odot}$. {\color{red}{Panel \emph{b} shows the magnitude and cz limits of ECO, where light grey dots indicate the $M_{r} < -17.33$ sample and black dots indicate our final approximately mass-limited sample with $M_{bary} > 10^{9.3} M_{\odot}$ and group membership within the ECO volume.}} }
\label{magsel}
\end{figure*}

\begin{figure}
\epsscale{1.1}
\plotone{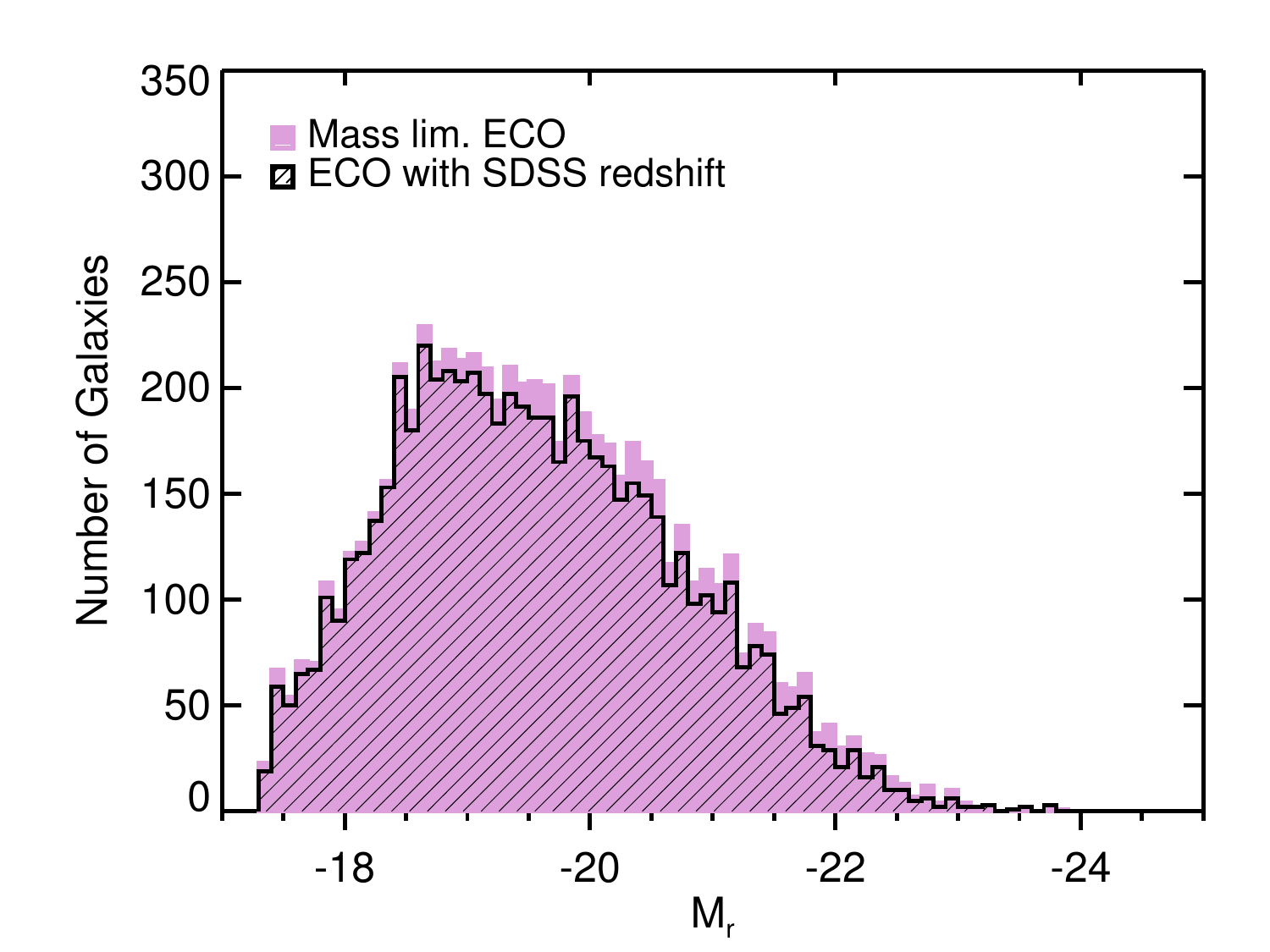}
\caption{Illustration of the additional completeness of the {\color{red}{final $M_{bary}$-limited}} ECO catalog over the SDSS redshift catalog. The $M_{r}$ distribution of the ECO sample in our reprocessed magnitude system (see \S \ref{photdesc} for details) is indicated by the purple solid histogram, and the distribution of ECO galaxies that have SDSS redshifts is indicated by the black hashed histogram.}
\label{magcz}
\end{figure}

We begin by introducing the galaxy samples under consideration in \S \ref{samp}. In \S \ref{allmeth}, we describe our main data analysis methods. In \S \ref{res}, we report a variety of results from this analysis, including the observed forms of both the traditional and alternative morphology-environment relations, in particular finding that the morphology-environment relation disappears for low baryonic mass central galaxies. We also find that blue-sequence early type, gas-dominated, and disk-growing populations rise in prominence in environments with $M_{halo} \lesssim 10^{11.5} M_{\odot}$. In \S \ref{disc}, we show that the forms of both the traditional and alternative morphology-environment relations we observe are consistent with expectations of the disk regrowth model and discuss the idea that the low group halo mass regime below $\sim10^{11.5}M_{\odot}$ appears to be a preferred regime for disk growth. Finally, we provide a brief summary of our major results in \S \ref{conc}.

\section{Samples}
\label{samp}

\subsection{ECO catalog}   
\label{ECOsamp}
  The ECO, or Environmental COntext, catalog is the largest sample we consider and includes the greatest diversity of galaxy environments (see Table 1 for detailed ECO galaxy properties). The ECO catalog region was chosen as the largest contiguous region on-sky where the highly complete Updated Zwicky Catalog (UZC; \citealp{UZCcat}) and Sloan Digital Sky Survey (SDSS; \citealp{SDSScat}) redshift databases overlap, allowing objects not present in either one to be recovered through inclusion of the other, with SDSS typically providing redshifts for fainter objects than the UZC. Though defined by the overlap of these two catalogs, the ECO catalog also incorporates redshifts from the REsolved Spectroscopy Of a Local VolumE (RESOLVE; Kannappan et al., in prep.), HyperLEDA \citep{Hyperledacat}, GAMA \citep{GAMAcat}, 2dF \citep{2dFcat}, and 6dF \citep{6dFcat} surveys. The ECO region was also selected to enclose the RESOLVE A-semester survey volume plus a minimum 1~Mpc ``buffer'' in all directions (see sky coverage in Fig.\ \ref{skycov}). This buffer region, chosen with a size comparable to typical {\color{red}{group}} halo virial radii at the present epoch, exists to mitigate potential edge effects in calculating galaxy environment metrics, such that only galaxies in the buffer region should have environmental measures strongly affected by the loss of nearby galaxies outside the catalog boundaries. The far side limit of 7470 km/s was selected to encompass both the aforementioned 1~Mpc (equivalent to 70 km/s for $H_{0} = 70~km~s^{-1}~Mpc^{-1}$) buffer beyond the RESOLVE cz limit of 7000 km/s and an additional allowance to compensate for group peculiar velocities up to 400 km/s. The near side buffer zone cz limit of 2530 km/s was similarly chosen to expand the ECO volume as much as possible beyond the near-side RESOLVE cz limit of 4500 km/s while avoiding the effects of Virgo Cluster region velocity-space distortions (see Fig.\ \ref{racz}). We consider the velocity limits of the non-buffer ECO volume to be 470 km/s away from the buffer edges, or 3000 km/s $< cz <$ 7000 km/s. 

The ECO catalog represents a cross match between sources with measured redshifts found in the UZC, SDSS (including data releases 6, 7, and 8; \citealp{SDSSdr6cat}; \citealp{SDSSdr7cat}; \citealp{SDSSdr8cat}), HyperLEDA, RESOLVE, GAMA, 2dF, and 6dF catalogs with a 15" matching radius on sky. New sources are added to ECO from each of the constituent catalogs whenever they do not match to a previously included ECO source. The resulting catalog has also been inspected by eye for duplicate entries caused either by ``shredding'' of SDSS photometric objects into multiple galaxy pieces (as described in \citealp{SDSSdr2cat}) or by centering/coordinate errors occasionally larger than the cross-matching radius. Such duplicate entries, making up $\sim$5\% of the galaxies originally considered for inclusion, have been removed from our catalog. As illustrated in Fig.\ \ref{magcz}, the majority of ECO galaxies are present in the SDSS redshift survey, but the number of galaxies added from other sources is significant, at approximately 7\% of the final catalog. 

Initially, sources with measured positions and redshifts inside the ECO volume are considered for potential membership regardless of any previously measured catalog magnitudes they may or may not have. We then use custom photometric measurements performed on SDSS imaging frames for all such potential ECO members (see \S \ref{photdesc} and \citealp{Eckert2015}) to determine a defining magnitude limit for the ECO catalog. The completeness limit of the SDSS redshift survey at 7000 km/s is $M_{r} = -17.23$ (in DR7 catalog Petrosian magnitudes corrected for foreground extinction), and our reprocessed magnitudes are typically brighter than the SDSS catalog values by approximately 0.1 mag (see \S \ref{photdesc} for details). Thus, a potential completeness limit for the ECO catalog motivated by the SDSS completeness would be $M_{r} < -17.33$, which we use to produce an initial group catalog as described in \S \ref{gfsec}. However, we seek to create a baryonic mass limited final sample for our analysis, where baryonic mass ($M_{bary}$) is defined here as stellar plus atomic gas mass and estimated as described in \S \ref{massmod} and \ref{HIest}. As illustrated in Fig.\ \ref{magsel}a, within the confines of the $M_{r} = -17.33$ magnitude completeness limit, we can construct an approximately baryonic mass limited sample with $M_{bary} > 10^{9.3} M_{\odot}$ while leaving aside only a relatively small number of high mass-to-light ratio objects.

The final ECO sample we analyze meets both the above limit and the additional criterion that the center of the \emph{group} to which each galaxy belongs (see \S \ref{envmet} for details of group membership determination) must lie within the limits of the non-buffer region of ECO, that is, the group center must have 3000 km/s $< cz <$ 7000 km/s and RA/Dec $>1$Mpc from the buffer edges on sky at its redshift (see Figs.\ \ref{racz} and \ref{magsel}b)\footnote{Note that we do not recalculate galaxy $M_{r}$ values by considering galaxies to lie at their group center redshifts. We find this recalculation would make a negligible difference in our overall sample membership.}. This final sample contains 6716 galaxies.

\begin{figure}
\epsscale{1.1}
\plotone{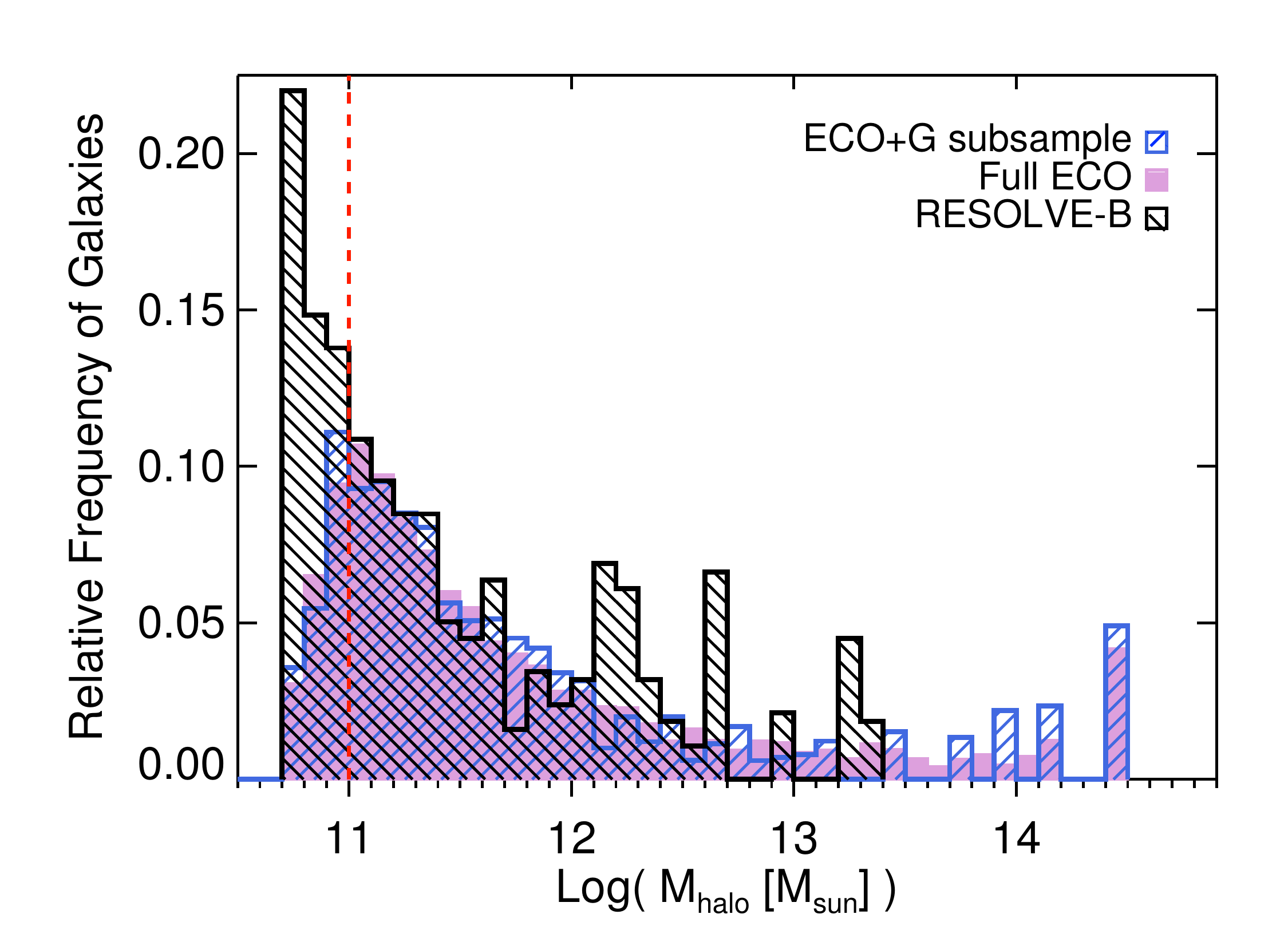}
\caption{The distribution of galaxies by {\color{red}{group halo mass in the mass-limited ECO, ECO+G, and RESOLVE-B samples, with completeness correction factors applied as described in \S \ref{ccs}.}} The red dashed vertical line indicates the group halo mass completeness limit for ECO, and the RESOLVE-B frequency histogram has been rescaled to match ECO at this limit. We have selected the ECO+G sample to have a similar environment distribution to the full ECO sample.}
\label{galexregmhalo}
\end{figure}

Two partially overlapping subregions of the ECO catalog are given special attention in this paper. The ``ECO+A'' region is defined by the overlap of the ECO catalog and public Arecibo Legacy Fast ALFA (ALFALFA) $\alpha.40$ catalog \citep{A40cat} and represents the portion of the ECO sample for which direct HI mass determinations are available and used in our analysis (see Fig.\ \ref{skycov}). The ECO+A region encompasses the full range of environments found in the ECO sample. The ``ECO+G'' subsample is defined by the availability of archival \emph{GALEX} imaging (\citealp{GALEX}; \emph{GALEX} MAST GR6/7 archive at \verb1http://galex.stsci.edu/GR6/1) with exposure times $>$1000s in the NUV band, sufficient to detect extended UV disk structures (e.g., \citealp{Thilker07}). Since \emph{GALEX} imaging coverage over the ECO sky region is patchy, we select multiple fully covered subregions of ECO as our final ECO+G subsample, largely coincident with RESOLVE-A but extending to larger Dec and including several slices through rich clusters. Together these regions closely reflect the full ECO environment distribution (see Fig. \ref{galexregmhalo}).

\subsection{RESOLVE-B} 
\label{Rfall}

  The B-semester region of the RESOLVE survey, which covers most of the SDSS ``Stripe 82'' region, is used as a comparison sample in this analysis. The RESOLVE-B region environment distribution is illustrated in Fig.\ \ref{galexregmhalo}, where primary differences compared to ECO are the lack of $M_{halo} \gtrsim 10^{13.5} M_{\odot}$ groups while $M_{halo} \sim 10^{13} M_{\odot}$ groups are overrepresented in RESOLVE-B. The RESOLVE-B subsample has the advantage of greater completeness than the ECO catalog, due to deeper SDSS imaging and redshift coverage, plus the further redshift completion efforts of the RESOLVE survey. By comparison with this extra-complete sample, we assess the effects of incompleteness in the ECO sample and derive completeness corrections that can be applied to ECO (median correction $\sim$1\%; see \S \ref{ccs}). For all galaxies in RESOLVE-B, morphological classification has been performed by a team of classifiers, providing both the basis for calibration of, and a comparison to, the quantitative morphological classifications used for ECO (see \S \ref{morphclass}; RESOLVE classifications details in Kannappan et al., in prep.). 

\begin{figure}
\epsscale{1.1}
\plotone{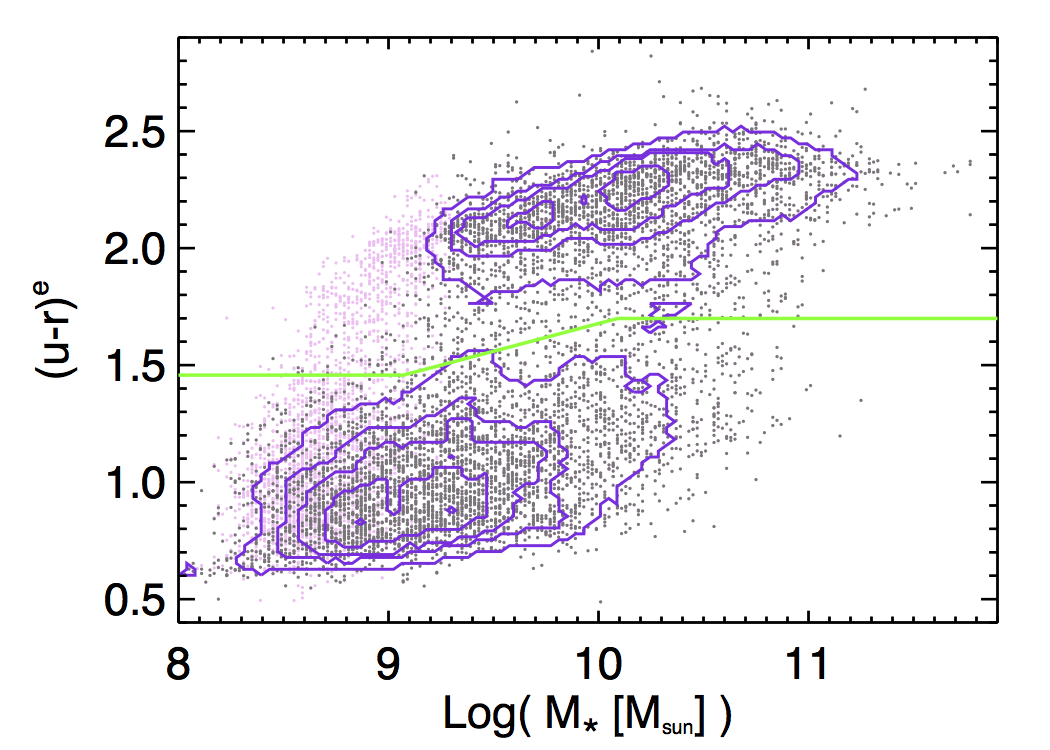}
\caption{Color vs.\ stellar mass for the ECO catalog sample, where $(u-r)^{e}$ represents an internal extinction corrected color derived from the SED fitting code (see \S \ref{massmod}). {\color{red}{Dark grey points indicate individual ECO galaxies in our mass-limited sample, light pink points indicate ECO galaxies that do not enter this final sample, and density contours of the mass-limited sample distribution are shown in purple.}} The green line indicates our chosen red/blue sequence divider (\S \ref{rbseq}).}
\label{Jcolmstar}
\end{figure}

\section{Methods}
\label{allmeth}

  In this section, we describe our methods of custom photometric processing, galaxy color and stellar mass estimation, galaxy morphology and UV disk classification, environment metric calculation, atomic gas mass estimation, and correction for the incompleteness of the ECO sample. Throughout our analysis, we calculate distances according to $D=cz/H_{0}$ and take $H_{0} = 70~km~s^{-1}~Mpc^{-1}$ unless otherwise noted. We estimate binomial confidence intervals on population proportions according to the Bayesian approach of \citet{Camconflims}.

\subsection{Imaging/Photometry}
\label{photdesc}
  As a result of the problematic nature of obtaining accurate estimates of galaxy properties with bulk SDSS pipeline-processed data and our desire to study spatially resolved parameters not necessarily computed in catalog data products, we have undertaken a custom reprocessing of SDSS, 2MASS, and \emph{GALEX} imaging for all ECO sample galaxies considered here. The ECO photometric reprocessing mimics the methods developed for the RESOLVE survey (see \citealp{Eckert2015} for full details).

To summarize this reprocessing: we retrieve imaging frames in $ugriz$, $JHK$, and NUV bands via automated queries to the SDSS DR8 \citep{SDSSdr8cat}, 2MASS \citep{2MASScat}, and \emph{GALEX} \citep{GALEX} archives respectively. Greater than 99\% of our galaxies are covered by 2MASS and $\sim$30\% by \emph{GALEX}. Photometric processing proceeds first on the SDSS imaging, where SExtractor \citep{sextractor} is called to identify sources from an $r$-band image and create a corresponding mask image wherein sources other than the target are masked. SExtractor parameters for the target galaxy are then used as an initial input to the IRAF \emph{ellipse} task, which fits the galaxy isophotes in a co-added $gri$ image while allowing the PA and ellipticity to vary. From this free ellipse fit, an optimal ellipticity and PA corresponding to the outer disk is determined. A fixed ellipse fit is then performed, using these outer-disk parameters, on the images in each band individually. Total magnitudes are determined from the resulting profiles by several methods: large aperture summation, exponential disk fitting, curve-of-growth extrapolation, and outer-disk color correction (see \citealp{Eckert2015}). Comparing the results of these methods yields an estimate of the systematic errors, which are combined with the purely photometric errors to obtain the final magnitude error estimates.

The automated SExtractor masking procedure has been tuned to give reasonable results for the majority of galaxies, but it is possible for the automatically generated masks to either mask parts of the galaxy under consideration or fail to mask nearby sources not associated with the target galaxy. To identify potentially problematic masks, we flag objects for which the magnitude estimation procedure has failed, the extracted $r$-band profile signal does not rise above eight times the sky noise, or the $r$-band profile does not extend beyond the calculated $r$-band 90\% light radius. These mask images flagged as potentially problematic are inspected by eye and edited by hand where necessary to better reflect distinctions between the target galaxy and other nearby sources. 

\begin{figure*}
\epsscale{1.1}
\plottwo{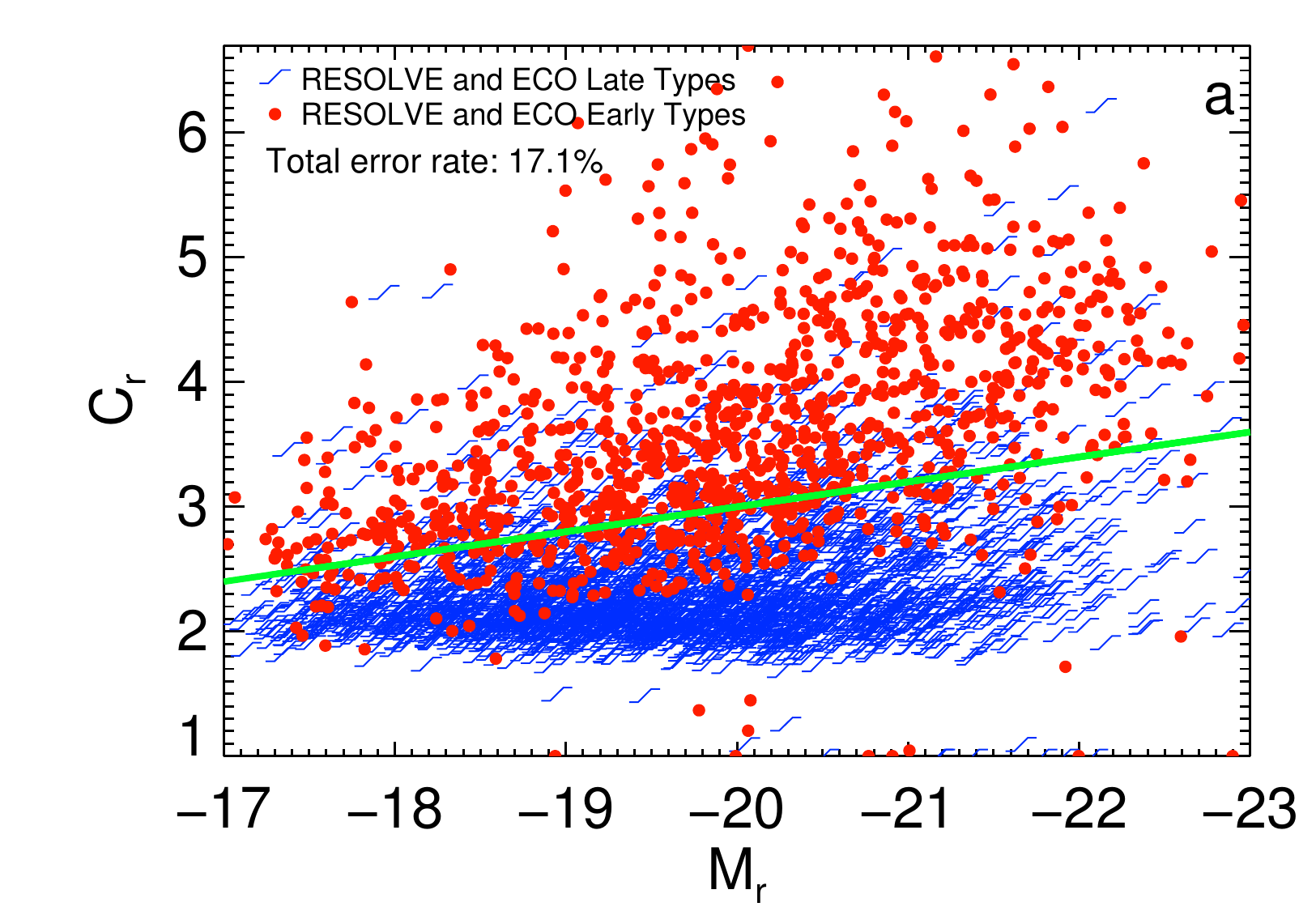}{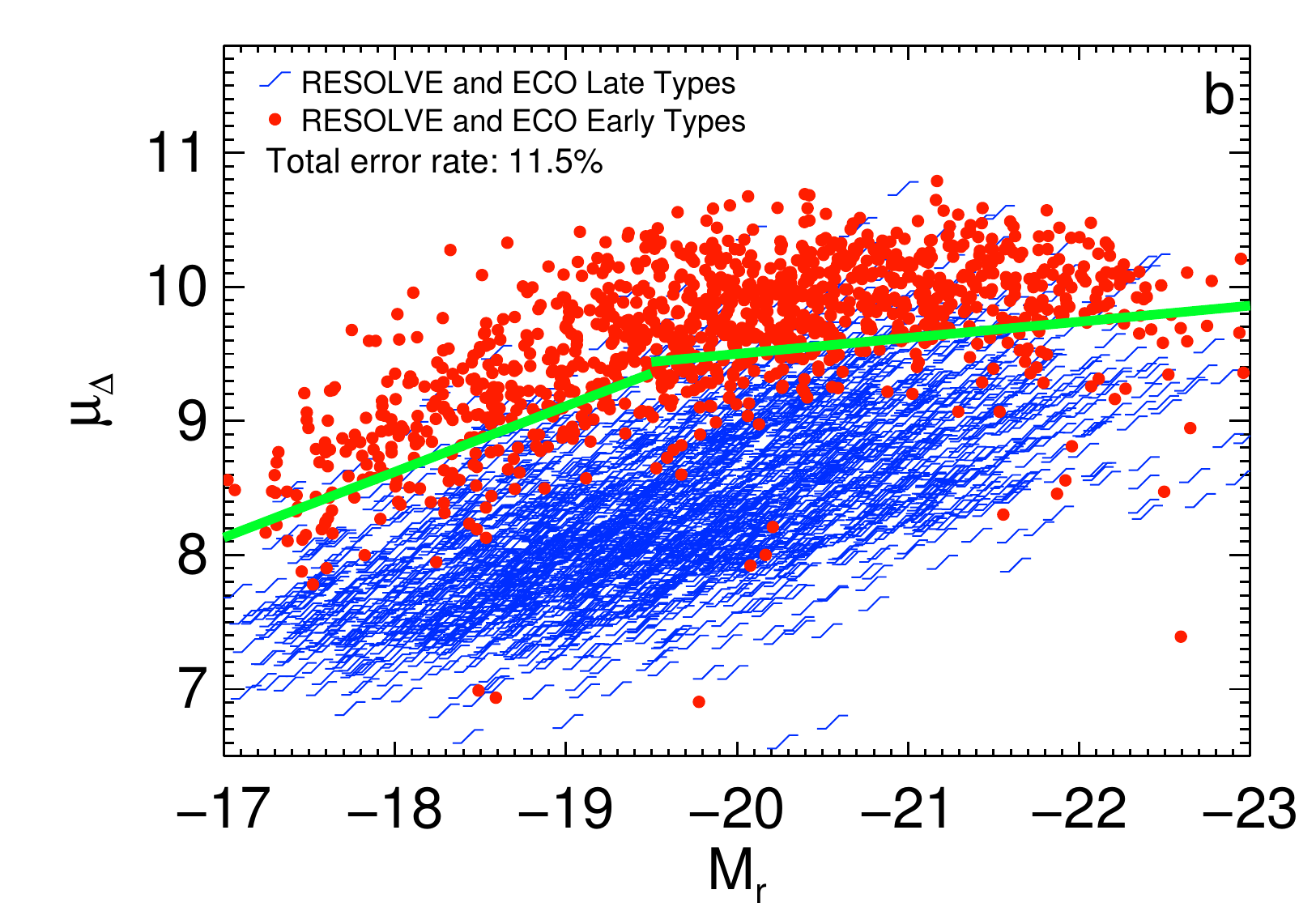}
\caption{Quantitative morphology metrics applied to by-eye classified galaxies in the RESOLVE and ECO samples (see \S \ref{morphclass} for details). Panel \emph{a} shows a cut in $C_{r}$ vs.\ $M_{r}$ applied to ECO and RESOLVE galaxies classified by eye as early and late types. The solid green line shows an optimized morphology discriminant in this parameter space ($C_{r} =-0.2 \times M_{r} - 1$), which performs poorly in duplicating the by-eye morphological classes. Panel \emph{b} shows the distribution of these same galaxies in the $\mu_{\Delta}$ vs.\ $M_{r}$ parameter space. The solid green lines show our optimized morphology discriminant in this parameter space, which gives improved classification error rates over the concentration index approach.}
\label{qmorph}
\end{figure*}

\subsection{Color and Stellar Mass Estimation}
\label{massmod}
  Using the full complement of total magnitudes, including NUV where available, galaxy stellar masses are estimated from a spectral energy distribution (SED) fitting procedure. We use a recently updated version of the likelihood-based stellar mass estimation code of \citet{KG07}, which is described fully by K13. This procedure uses a suite of composite stellar population models constructed from old and young \citet{BC03} stellar populations. These model stellar populations are combined in various fractions by mass, with 13 allowable young population fractions of 0.001, 0.002, 0.005, 0.011, 0.025, 0.053, 0.112, 0.220, 0.387, 0.585, 0.760, 0.876, and 0.941. Four possible metallicities (Z $=$ 0.004, 0.008, 0.02, and 0.05) and 11 possible extinction values ($\tau_{V}$ $=$ 0, 0.12, 0.24...1.2) are also used. The young population model grid is constructed to simulate both continuous and bursty star formation histories by including models with constant star formation from 1015 Myr in the past to various end points 0-195 Myr in the past and simple stellar populations (SSPs) with ages 360, 509, 641, 806, and 1015 Myr. The old population model grid includes SSPs with ages 2, 4, 6, 8, 10, and 12 Gyr. A Chabrier initial mass function \citep{Chabrier03} is used in these calculations, which yield a stellar mass zero point consistent with that of \citet{Kauffmann03ms}. The code also outputs internal extinction corrected model fit colors, which more cleanly separate the red- and blue-sequence galaxies than raw measured colors, and we designate these colors with a superscript ``e''. We also use model fit colors without any internal extinction correction, denoted with superscript ``model''.

\subsection{The Red and Blue Sequences}
\label{rbseq}
  To separate red- and blue-sequence galaxies, we choose a
dividing line between the red and blue color stellar mass loci defined by our extinction-corrected colors and stellar mass estimates (see Fig.\ \ref{Jcolmstar}). We determine this division based on double Gaussian fits to the red- and blue-sequence color distributions in two high and low stellar mass regimes where the sequences are well defined ($\log{M_{*}} < 9.5$ and $\log{M_{*}} > 9.5$). Our divider is then defined by the color halfway between the fit peaks and the median stellar mass in each mass regime. The slanted divider in the intermediate mass regime is defined by the line connecting these two Gaussian-fit-determined points. The equation of the divider is:
\begin{equation}
\resizebox{.85\hsize}{!}{%
$(u-r)^{e}=
\begin{dcases}
  1.457            & \log{M_{*}} \leq 9.1\\
  0.24 \times \log{M_{*}}-0.7 & 9.1 < \log{M_{*}} < 10.1 \\
  1.7           & \log{M_{*}} \geq 10.1 \text{.} \\
\end{dcases} $
}
\end{equation}

\subsection{Morphology Classification}
\label{morphclass}

  To calibrate a quantitative morphology cut for application to the ECO galaxies, we use by-eye morphological classifications from the RESOLVE survey, the A-semester sample of which is largely a subset of the ECO catalog (see Kannappan et al., in prep.\ for full details). RESOLVE galaxies that were given uncertain classifications by the classifiers are omitted from consideration. Since this comparison sample has just over 1000 galaxies, considerably fewer than the full ECO sample, and possesses few very bright galaxies, we also add to our morphology calibration sample those galaxies in ECO that have been previously classified by eye in the catalog of \citet{Naircat} or by the Galaxy Zoo Project \citep{Galzooclass}. We use only the ``clean and debiased'' Galaxy Zoo classifications referenced by \citet{Galzooclass}, which require 80\% of classifiers to agree on the chosen morphological type and debiasing with respect to the effects of luminosity, size, and distance on the classifications (see \S 3.1 of \citealp{Galzooclass}).

Based on comparisons to this by-eye classified sample, an optimal quantitative morphology cut was derived for application to ECO. Traditional quantitative morphology discriminants, such as the concentration index $C_{r} = R_{90\%}/R_{50\%}$ defined using SDSS catalog photometry (e.g., \citealp{Strateva01}; \citealp{Shimasaku01}), yield unfortunately high error rates in the ECO sample (see Fig.\ \ref{qmorph}a)\footnote{Note that we explicitly avoid the inclusion of color as a parameter used for morphology discrimination due to the bias it would introduce against blue early and red late types.}. We instead employ the $\mu_{\Delta}$ metric recently developed by K13, which combines the surface mass density within $R_{90\%}$ and a multiple of 1.7 times the difference between the surface mass densities within $R_{50\%}$ and within the $R_{50\%}-R_{90\%}$ annulus. We optimize this metric for use as an early/late type discriminant in ECO by considering it as a function of $M_{r}$ and choosing two linear cuts in this parameter space that yields the minimum misclassification rate in two separate $M_{r}$ regimes (see Fig.\ \ref{qmorph}b). Our optimized morphology discriminant is defined by:
\begin{equation}
\resizebox{.85\hsize}{!}{%
$\mu_{\Delta}=
\begin{dcases}
  -0.12 \times M_{r} + 7.1 & M_{r} \leq -19.5 \\
  -0.49 \times M_{r} - 0.2 & M_{r} > -19.5 \text{.} \\
\end{dcases} $
}
\end{equation}

To reduce the misclassification rate resulting from implementing a quantitative morphology cut, we make further use of the Galaxy Zoo clean and debiased morphology classifications. Where such classifications exist for ECO galaxies, we use the Galaxy Zoo early/late type classification rather than that inferred from the optimized quantitative morphology cut described above. This substitution results in a naive apparent misclassification rate of $\sim$3.19\% for late types and $\sim$15.0\% for early types in the morphology calibration sample.

We can better estimate the misclassification errors that would result from applying this quantitative calibration to an independent sample with a bootstrap resampling approach applied to the calibration sample, using the ``.632'' error rate estimator as described by \citet{Efron83}. The bootstrap sampling procedure is repeated for 1000 iterations. For each iteration, we randomly sample N objects from the morphology calibration sample with replacement, where N is equal to the total number of objects in the calibration sample (the ``.632'' nomenclature refers to the fact that $\sim$63.2\% of the objects will end up in each bootstrap sample on average because objects can be selected twice). We then use each bootstrap sample to determine an optimum classification rule and evaluate the misclassification rate using this rule when applied to those members of the morphology calibration sample that were outside the bootstrap sample. As for the ECO sample, we include replacement of quantitatively inferred classifications with those from Galaxy Zoo to fairly assess the error rate during each iteration. We calculate our final error estimates as:
\begin{equation}
E_{final}=0.632E_{bavg}+0.368E_{app},
\end{equation}
where $E_{bavg}$ represents the average of the estimated error rates over the 1000 resampling iterations and $E_{app}$ represents the naive apparent error rate estimate in the calibration sample. The final error estimates are similar to the aforementioned naive apparent error rates, with estimated $\sim$3.23\% error rate for late types or $\sim$15.8\% for early types. We also calculate these error estimates in multiple individual bins of group halo mass and galaxy stellar mass as the error rate is not necessarily constant for different subclasses of galaxies in the sample. Our early-type misclassification rates tend to decrease with increasing galaxy stellar mass, from $\sim$19\% at our lowest stellar masses to $\sim$11\% at our highest stellar masses, and to increase with increasing group halo mass, from $\sim$14\% at our lowest halo masses to $\sim$17\% at our highest halo masses.

\begin{figure}
\epsscale{1.15}
\plotone{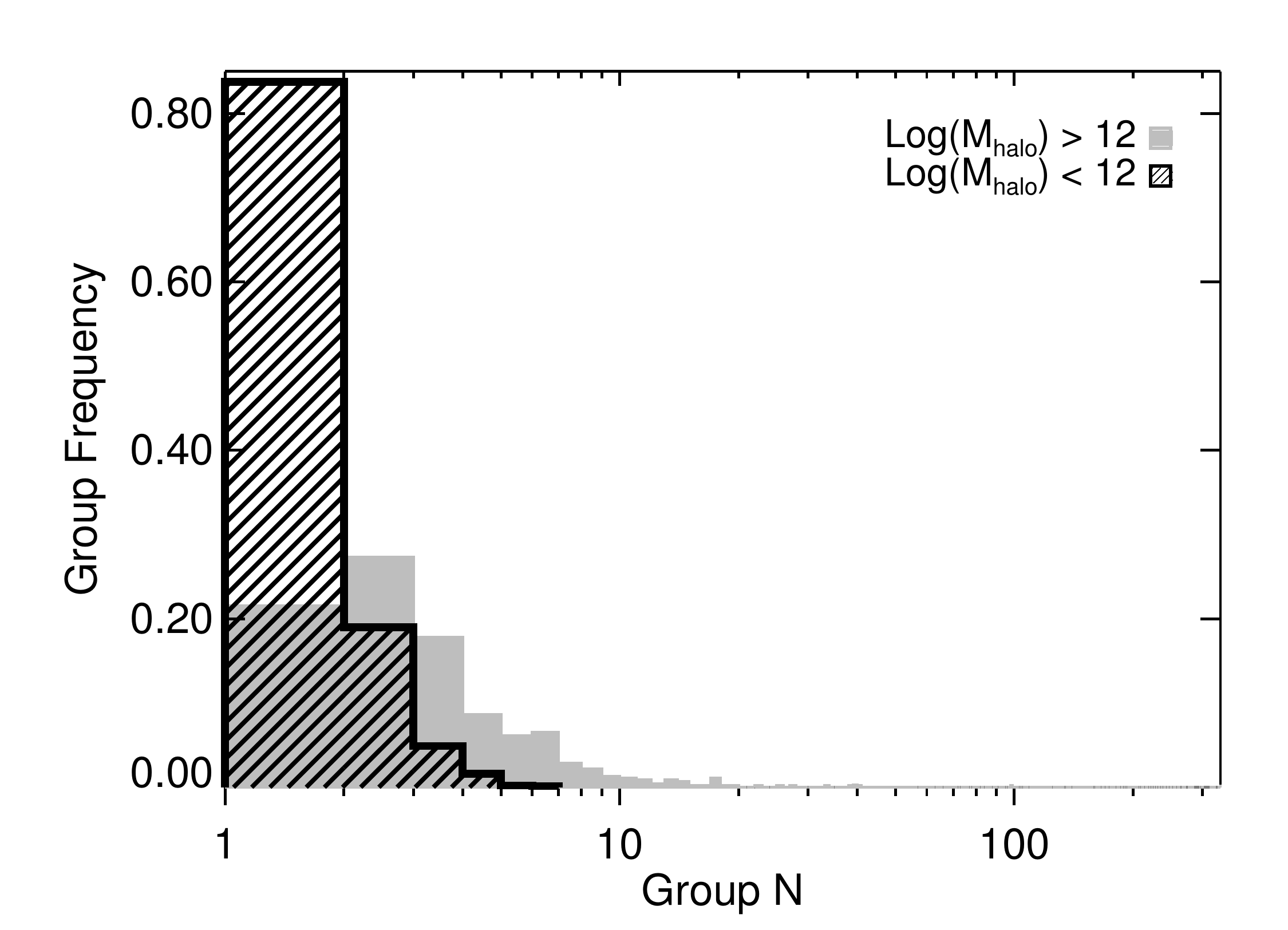}
\caption{Illustration of the distributions of group membership that correspond to low and high estimated group halo mass regimes in our group finding analysis.}
\label{groupnhist}
\end{figure}

\begin{figure}
\epsscale{1.1}
\plotone{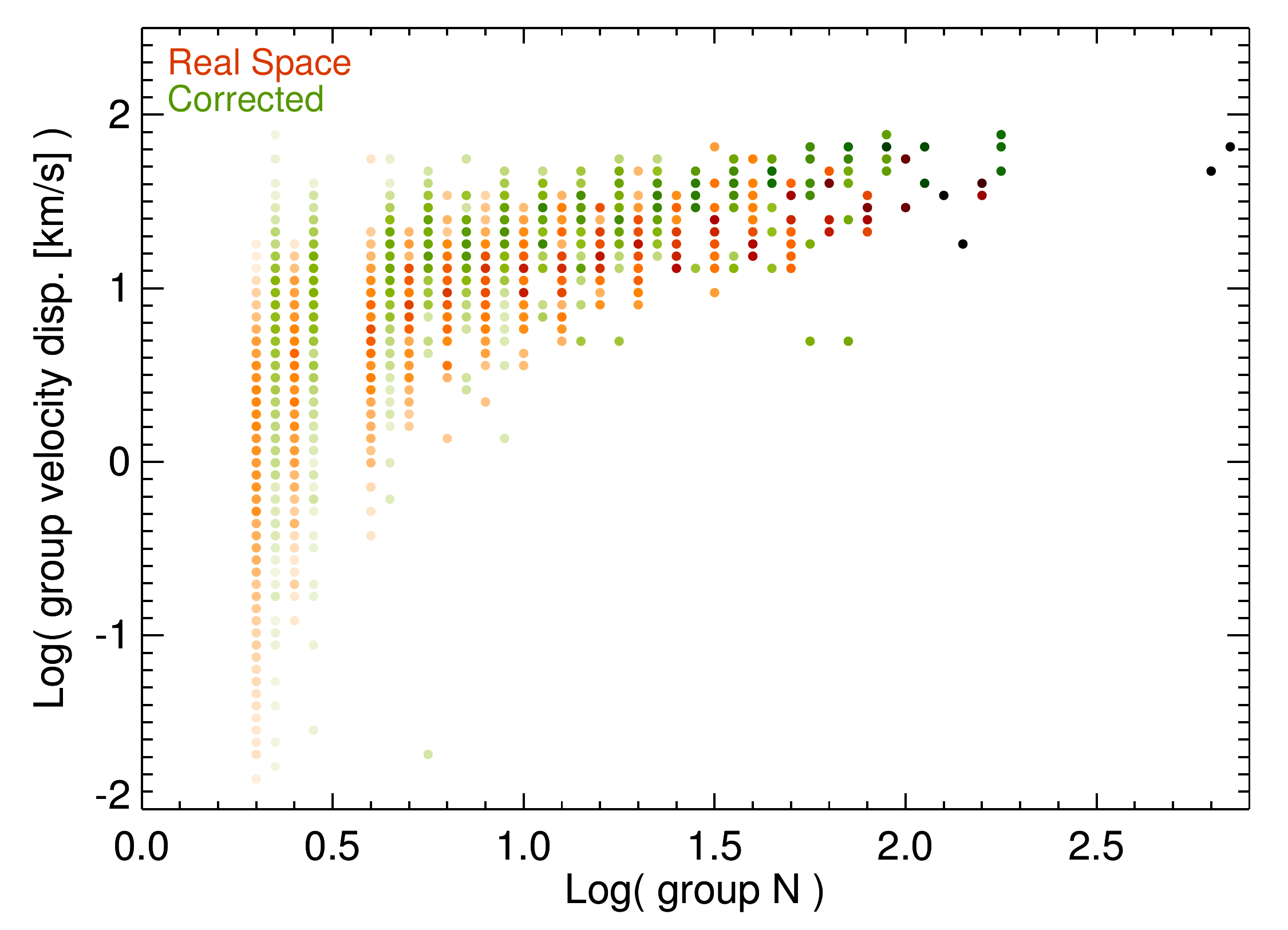}
\caption{Illustration of the recovery of real-space mock galaxy catalog group velocity dispersions after running group finding on a redshift-distorted version of the catalog and applying our finger-of-God collapse procedure on the resulting group catalog. Plotted are the distributions of real-space group velocity dispersion (red/orange scale) and corrected group velocity dispersion (green scale) versus real group N, where color coding of points represents the histogram density normalized to sum to one within each bin of group N (darkest colors imply the highest density). Red and green points are plotted offset by half the group N bin size for clarity. Note that we typically do not collapse groups to velocity dispersion levels as small as those seen in real-space groups.}
\label{fogvalid}
\end{figure}

\subsection{Identification of UV Disks}
\label{UVids}

For the ECO galaxies covered by \emph{GALEX} NUV imaging, we apply an automated procedure for quantitatively identifying Ultraviolet-Bright (UV-B) disks according to the definition of \citet{Me12}. For each galaxy, we apply the center position, position angle, and ellipticity of the SDSS optical ellipse fit to a set of fixed parameter IRAF \emph{ellipse} fits on the NUV imaging, which are allowed to proceed outwards radially until no significant UV flux is detected. We apply additional SExtractor-derived masking of the UV images beyond that determined for the original SDSS optical photometry procedure, necessary due to the low resolution of the \emph{GALEX} data and the occasional appearance of new contaminating sources not present in the optical data. As described by \citet{Me12}, the quantitative UV-B disk classification we employ requires satisfaction of an NUV$-K$ color condition, which is included to ensure young stellar population ages and select stellar populations with a minimum $\sim$10\% young component by mass. We calculate NUV$-K$ colors between the optical $g$-band 50\% light radius and the end of the NUV profile and require NUV$-K$ $<$ 4.5 for classification as a UV-B disk\footnote{Note that in some cases galaxy outer disks are not well detected in our 2MASS images. In such cases, we calculate upper limit K-band magnitudes in the outer-disk regions and use these magnitudes for determining the NUV$-K$ color.}.

Validating this fully automated identification approach against the methods of \citet{Me12}, which employed a more detailed, galaxy-by-galaxy elliptical isophote fitting procedure, our new algorithm identifies up to $\sim$25\% more UV-B disks in an E/S0 sample. Reasons for the difference between the more and less automated approaches include occasional position angle misalignments between the galaxy regions where optical and UV emission dominate and imperfect automated masking of UV sources, which is in some cases less aggressive than the masking employed by \citet{Me12}.

\subsection{Environment Metrics}
\label{envmet}

  For the catalog samples under consideration, we compute two different metrics of galaxy environment. We primarily focus on environmental trends using group halo mass, but we also compare to results obtained with a smoothed galaxy density field in \S \ref{res}. 

\begin{figure*}
\epsscale{1.}
\plotone{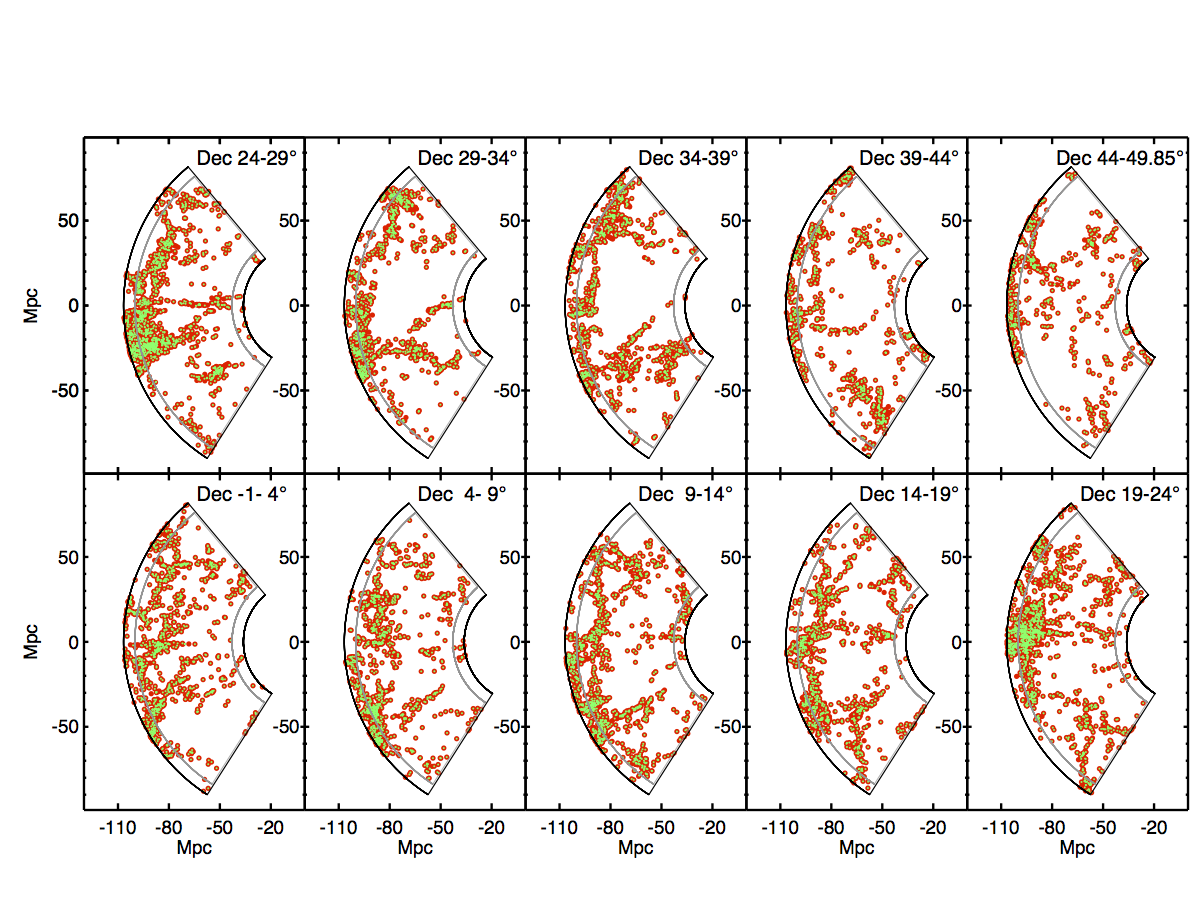}
\caption{Illustration of the finger-of-God collapse procedure applied to the ECO catalog. Red points show the original inferred spatial distribution of galaxies, and smaller green points show the distribution after the finger-of-God collapse procedure has been applied.}
\label{3dcollapse}
\end{figure*}

\subsubsection{Group Finding}

\label{gfsec}
We identify groups of galaxies using the friends-of-friends algorithm of \citet{fofalgo}. We infer group halo masses based on the observed total $r$-band luminosity of galaxies in each group following an abundance matching procedure, as described in \citet{BB07}. {\color{red}{The calculation of total luminosities for each group relies on high completeness to a fixed luminosity floor, so our group catalog is constructed for all ECO galaxies with $M_{r} < -17.33$, even though we further analyze only a subsample of these galaxies satisfying our $M_{bary}$ limit. Our group halo mass estimates assume a monotonic relation between the observed group luminosity and its halo mass. This relationship is determined by matching the space density of observed groups of a given luminosity to the theoretical space density of dark matter halos of a given mass, as derived from a standard concordance cosmology halo mass function.}} From consideration of the mock galaxy catalogs described below, we find that typical group halo mass errors are of order 0.12 dex, although much larger errors can sometimes occur when groups are improperly fragmented or improperly linked together.\footnote{Note that the zero point of our group halo mass scale is dependent on comparisons with the mock catalog, as we calibrate our group halo masses to best approximate the true group halo masses in the mock catalog for the majority of halos in our sample, i.e., those at low mass. We subtract a simple constant correction of $\sim$0.15 dex from our initial abundance-matched mass estimates to correct for a tendency towards overestimation at low group halo mass. A significantly more complicated correction would be required to yield perfect reproduction of the true group masses in all mass regimes, but using this simple correction we obtain median estimated-mass-minus-true-mass offsets of only 0.02 dex for $M_{halo} < 10^{12} M_{\odot}$ and -0.12 dex for $M_{halo} > 10^{12} M_{\odot}$, which are less than or consistent with the typical group halo mass errors overall.} {\color{red}{We label the brightest galaxy in each group the central and all other galaxies as satellites, and imperfect group identifications can affect these central vs. satellite designations, adding noise to comparisons of central vs. satellite trends. Note that in some cases the galaxy collections we refer to here as ``groups'' can in fact consist of single detected galaxies (that is, groups with N=1), which occurs most often for the lowest mass ``groups'' we probe (see Fig. \ref{groupnhist}).}}

The group-finding algorithm we employ automatically determines an appropriate ``linking length'' for grouping individual objects together, {\color{red}{equal to $b_{\perp}$ and $b_{\parallel}$ times the mean separation between galaxies in the input sample in the on-sky and line-of-sight directions, respectively (with $D=cz/H_{0}$ distances and $H_{0} = 100h~km~s^{-1}~Mpc^{-1}$ used for this calculation). Following the recent analysis of \citet{DM2014} we use values of $b_{\perp}=0.07$ and $b_{\parallel}=1.1$ for our group finding, as these values are found to be optimal for creating group catalogs that will be used for statistical study of the effects of galaxy environment.}} For a sample region as small as RESOLVE-B, this mean spacing determination is far more sensitive to cosmic variance than for a large region as in the ECO sample. To compensate, when we apply the group-finding algorithm to RESOLVE-B we fix the linking lengths to those determined from a version of the ECO catalog limited to $M_{r} < -17$, the magnitude limit of RESOLVE-B. We also use the luminosity to {\color{red}{group}} halo mass conversion determined from this version of the ECO group catalog, since this conversion is similarly sensitive to cosmic variance. Through testing with an ``ECO-analog'' version of RESOLVE-B (matched to ECO's shallower depth and lower completeness; see \S \ref{ccs}), we find that the difference in completeness between RESOLVE-B and ECO does not significantly bias the group halo masses we estimate, with scatter between the group halo masses inferred from the two versions of RESOLVE-B only reaching $\sim$0.2 dex at $M_{halo} \gtrsim 10^{13} M_{\odot}$. The primary effect of extra completeness on group identifications is that additional $N=1$ halos, typically containing faint galaxies near the magnitude limit, are found. Fig.\ \ref{galexregmhalo} reveals that ECO is not complete for group halos with masses less than $M_{halo} \sim 10^{11} M_{\odot}$, so we refrain from including {\color{red}{group}} halos below this mass limit in our analysis.

\subsubsection{Smoothed Galaxy Density Field}

  We calculate a smoothed galaxy density field with an IDL procedure based on the approach of \citet{smthdalgo}. This procedure takes in individual galaxy redshift-space positions, assuming line of sight $D=cz/H_{0}$ for consistency with other methods applied here, and creates a continuous number density field from this spatial point distribution by smoothing each galaxy with a unit-normalized Gaussian kernel and summing the resulting galaxy space and velocity distributions. Since the samples considered in this work are volume- and not flux-limited, no luminosity function weighting factors have been applied in the density field calculation (cf.\ \citealp{smthdalgo}). 

In using this smoothed galaxy density field procedure, we use a smoothing kernel width of $\sim$1.43~Mpc (1~Mpc/h with $h = 0.7$), a scale that is similar to a typical {\color{red}{group}} halo virial radius at $z\sim$0. In all subsequent density analysis, we limit ourselves to consideration of galaxies that lie $>$2 smoothing lengths from catalog edges and report density values normalized by the median of the density field values associated with these galaxies.

In order for densities smoothed on this small scale to be physically meaningful, we also implement a procedure to statistically collapse ``fingers of God'' in ECO, as these cause measured line-of-sight velocities in large groups/clusters to imperfectly reflect physical distances. Similar methods for statistically correcting redshift-space distortions have been applied by other authors (e.g., \citealp{Tegmark04}). The algorithm we apply was developed and calibrated on mock galaxy catalogs derived from an N-body simulation of a $\Lambda$CDM cosmological model, with $\Omega_{m}=$0.25, $\Omega_{\Lambda}=$0.75, $\Omega_{b}=$0.04, h$=$0.7, n$_{s}$=1.0, and $\sigma_{8}=$0.8. Initial conditions were set using second order Lagrangian Perturbation Theory \citep{2LPT} at a starting redshift of $z=$99 and the particle distribution was evolved using the code GADGET-2 \citep{Springel05}. The simulation contains 1050$^{3}$ dark matter particles in a box of size 180 Mpc/h, sufficient to encompass multiple ECO catalog volumes. The resulting particle mass is 3.5$\times10^{8}~M_{\odot}$/h and the gravitational force softening is 7 kpc/h. At this resolution, the lowest mass halos in the ECO catalog sample typically contain ~160 particles. We identify halos in the dark matter particle distribution using a friends-of-friends algorithm with a linking length equal to 0.2 times the mean interparticle separation. We then populate these halos with galaxies using a halo occupation distribution \citep{BW02} designed to produce a galaxy population with similar space density and clustering properties as ECO galaxies. Central galaxies are given positions and velocities of halo centers of mass, and satellite galaxies are given positions and velocities of randomly selected dark matter halo particles. This procedure produces a ``real-space'' mock galaxy catalog. The real-space catalog is subsequently distorted into a "redshift-space"  mock galaxy catalog by assuming the line of sight direction to extend radially outward from the center of the box and incorporating galaxy peculiar velocity components along the radial direction into galaxy redshifts.

Comparisons between real-space and redshift-space versions of the mock galaxy catalogs allow us to determine the distribution of 3D, real space radius offsets from the group center positions as a function of the group virial radii ($R_{vir}$), with $R_{vir}$ estimated using the group masses and the defining group overdensity. The overall real-space group-centric $R/R_{vir}$ density profile determined from the mock catalogs is approximated with a gamma probability density function fit and used as an input when we apply the collapse algorithm on observed catalogs. The fit relation has the following form:
\begin{equation}
f(x,a,b) = \frac{x^{a-1} \times e^{-bx} \times b^{a}}{\Gamma(a)} \text{,} 
\end{equation}
where a$=$1.6 and b$=$5.

We begin the finger-of-God collapse process on the observed galaxy catalogs with group identifications determined as described in \S \ref{gfsec}. For each galaxy in an $N>1$ group, we first assign a random real-space displacement from the center of the group according to our fit distribution of mock-catalog-determined $R/R_{vir}$ values, taking into account the group's estimated $R_{vir}$. Each assigned 3D radius is then checked for consistency with the galaxy's observed projected distance from the group center, that is, a galaxy's assigned 3D distance from the group center is required to be greater than its observed projected distance from the center. If this is not the case on first assignment, a different random 3D radius value is chosen until the condition is met. This procedure does not significantly bias the assigned $R/R_{vir}$ distribution. Next, the appropriate redshift direction displacement from the group center is determined for each galaxy, using the observed spatial coordinates and the assigned 3D radius. The sign of the redshift displacement for each galaxy relative to the group center is determined randomly.

Validation of the final collapse procedure on the mock catalogs yields reasonable agreement between the corrected and original simulated, real-space group velocity dispersions (see Fig.\ \ref{fogvalid}), although we tend not to reach corrected group velocity dispersions as small as the original dispersions seen in real-space mock catalog groups). This bias appears to be due largely to imperfect recovery of {\color{red}{group}} halo $R_{vir}$ by our group finding method; overestimation of $R_{vir}$ has the effect that our collapse algorithm places galaxies at larger 3D group-centric radii (and larger velocities) than observed in the real-space groups. The effect of the collapse process on the distribution of galaxies in the ECO catalog is shown in Fig.\ \ref{3dcollapse}. The application of this collapse process yields greater confidence in our density field results, which generally agree with group-finding results that are less sensitive to such finger-of-God effects.

\begin{figure*}
\epsscale{1.15}
\plotone{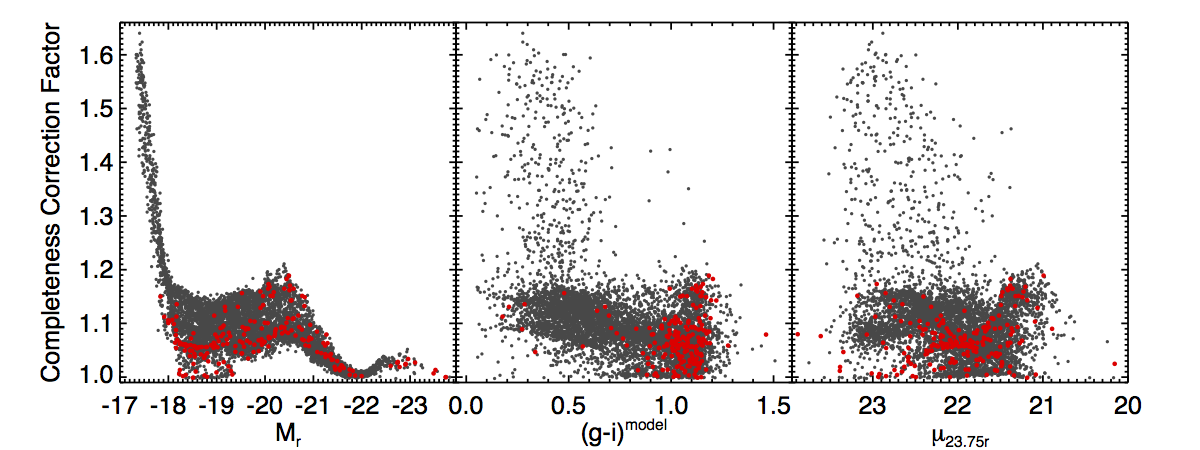}
\caption{Multiplicative completeness correction factors derived for each individual ECO {\color{red}{mass-limited sample}} galaxy (points) as a function of $r$-band absolute magnitude, model $g-i$ color, and {\color{red}{$r$-band surface brightness $\mu_{23.75r}$}}. Galaxies indicated by red points require further multiplicative correction factors on top of the factors shown, due to the loss of galaxies with extreme line-of-sight velocities outside the ECO definition. The majority of galaxies indicated by red points lie in the Coma Cluster, where the additional multiplicative correction factor reaches its maximum value of $\sim1.42$.}
\label{ccfield}
\end{figure*}

\subsection{Atomic Gas Mass Estimates}
\label{HIest}
We cross-match to published ALFALFA $\alpha.40$ \citep{A40cat} HI sources to find possible ECO galaxy matches and check for possible confusion using a large combined redshift catalog containing ECO (see \S \ref{ECOsamp}). We flag a source as confused if more than one galaxy is found within the 3$'$ matching radius (1.5 times the effective ALFALFA beam radius) and if the HI profiles would overlap in velocity space assuming minimum 50 km/s redshift uncertainties and typical line widths of $\sim$100 km/s (the weighted average of the ALFALFA-derived velocity width function; \citealp{Papastergis11}). We use the reported line flux densities to convert to $M_{HI}$ via the standard formula ($M_{HI}=2.36 \times 10^{5} \times D^{2} \times F_{HI}~M_{\odot}$; \citealp{HG84}). We then multiply our masses by 1.4 to correct for the presence of He. In the absence of HI detections, we estimate upper limits according to the procedure described by K13, which uses the typical $\alpha.40$ rms noise as a function of declination integrated over a velocity interval estimated for each source according to the average relation between internal velocity and $r$-band magnitude ($\log{V = -0.29 - 0.123~M_{r}}$, calibrated by K13). 

Where ECO galaxies lack ALFALFA $\alpha.40$ detections or have confused HI measurements, we use the ``photometric gas fraction'' technique \citep{K04} to assign HI mass estimates based on the observed tight correspondence between HI gas to stellar mass ratio and $u-J$ color. We employ the photometric gas fraction calibration and procedure described by K13, in which gas masses are assigned to galaxies with $(u-J)^{model} < 3.7$ according to the relation: 
\begin{equation}
\log{(M_{HI}/M_{*})=2.7-0.98~(u-J)^{model}}
\end{equation}
with random 0.34 dex scatter motivated by the observed relation. If the gas mass estimated with this relation for a given galaxy would exceed its calculated upper limit, the upper limit is adopted instead. For galaxies with colors redder than $(u-J)^{model} = 3.7$, that is, where the linear color-gas-to-stellar mass ratio relation breaks down for quenched galaxies, the procedure assigns random values in the logarithmic range $M_{HI}/M_{*} = 0.001-0.5$, again constrained to lie below the estimated upper limit for each galaxy.

We make one significant modification to the K13 photometric gas fraction procedure, which is motivated by the tendency for the bluest and most gas-rich galaxies to lie above the aforementioned gas-to-stellar mass ratio versus color relation (see K13 Fig.\ 8a). To better reflect the typical gas-to-stellar mass ratios observed for galaxies with $(u-J)^{model} < 3$, we multiply the gas mass estimates that would result from the K13 estimator by an additional factor of 1.5 in this regime, with resulting estimates still constrained to lie below upper limits where available. While our gas mass estimates are improved by the inclusion of this factor, we note that its inclusion does not qualitatively affect the results we report, which remain similar even if the extra multiplicative factor is omitted from estimated gas masses. In particular, we largely focus our analysis of gas content on the incidence of gas-dominated ($M_{HI}/M_{*} > 1$) galaxies in ECO, and so for much of our analysis, the exact gas mass estimated for a given galaxy is less important than whether or not it falls into a broad gas-to-stellar mass ratio category.

\subsection{Completeness Corrections}
\label{ccs}

\begin{figure*}
\plotone{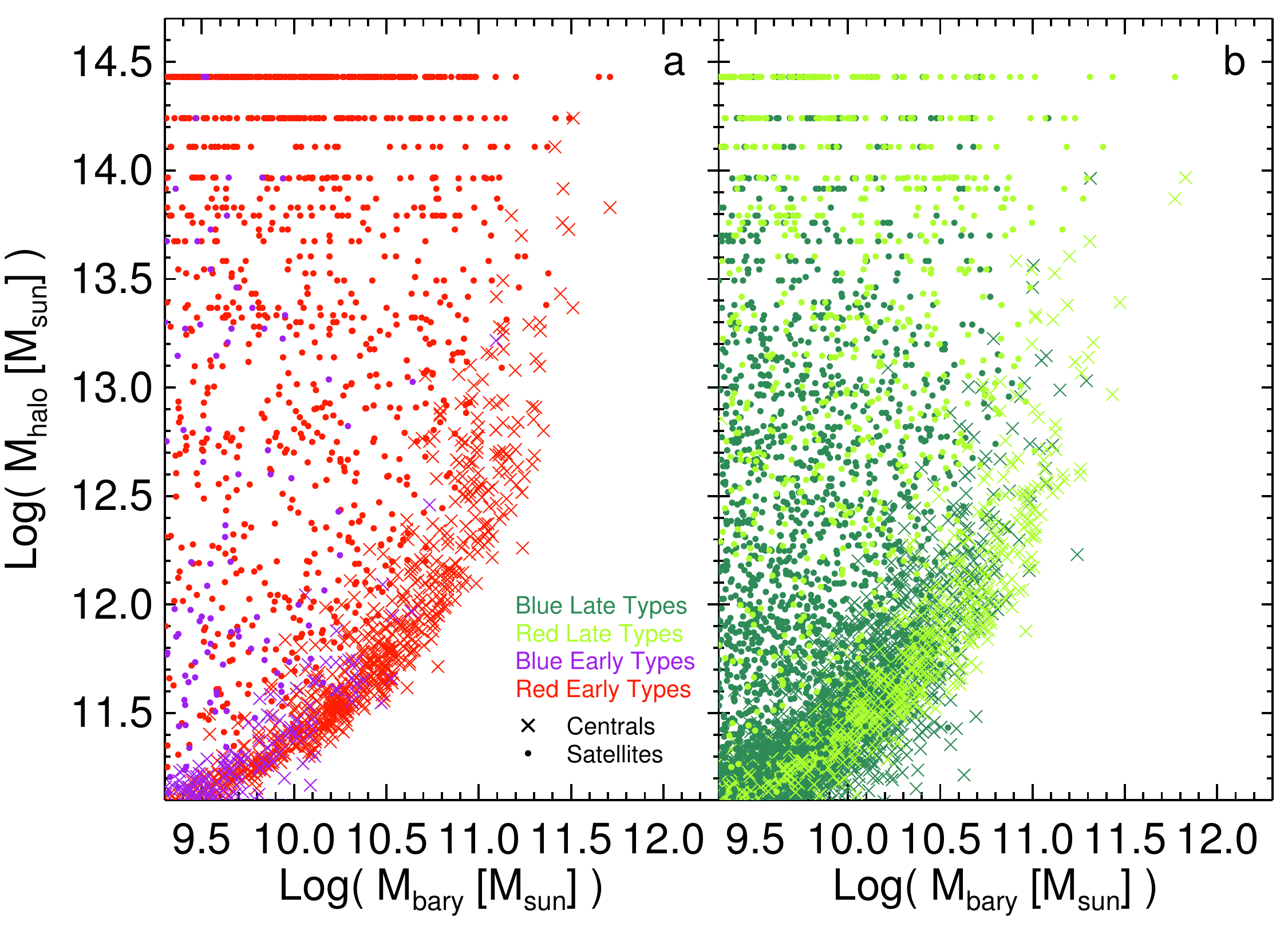}
\caption{Illustration of ECO catalog galaxy distributions in the group halo mass and galaxy baryonic mass space, where early types of red/blue and central/satellite classes are shown on panel \emph{a}, and late types of red/blue and central/satellite classes are shown on panel \emph{b}. No catalog membership completeness corrections have been applied in this figure.}
\label{mhalomstarscat}
\end{figure*}

  Since the RESOLVE-B sample has much higher completeness than the ECO sample (see \S \ref{Rfall}), we use the former to calibrate and correct for the effects of redshift incompleteness in the latter. {\color{red}{We do so in two steps: first we compare the full RESOLVE-B galaxy number counts with a version of RESOLVE-B that contains only galaxies with DR7 redshifts (that is excluding the extra SDSS Stripe 82 redshift coverage beyond the original redshift survey), and next we carry out a similar comparison between ECO catalog galaxy number counts and those of a version of ECO that contains only galaxies with DR7 redshifts. By dividing the RESOLVE-B DR7-to-full catalog correction factor by the ECO DR7-to-full catalog correction factor, we obtain a measure of the factor needed to correct the ECO catalog membership to the same completeness as the RESOLVE-B sample. For both steps of this process, we divide both samples into cells in two-dimensional parameter space grids (e.g., $M_{r}$ vs. color), using grids determined with a simple adaptive approach, which begins with subdividing each axis into four large cells. If more than a set minimum number of galaxies are present in a given cell (minimum is five for RESOLVE-B and 100 for ECO), the cell is subdivided in half iteratively until no further subdivisions are allowed by the minimum number of galaxies per bin condition. We interpolate each irregularly gridded dataset into a smoothly varying number density field in order to derive the final completeness correction fields. We have examined multiple different 2D parameter space options for deriving these corrections and find that in all cases the resulting corrections yield overall similar results. In order to reduce noise in the final correction caused by low-occupation bins, we choose to average the corrections derived for individual galaxies using the $M_{r}$ vs. $(g-i)^{model}$ and $M_{r}$ vs. $\mu_{23.75r}$ parameter spaces (where $\mu_{23.75r}$ represents an average $r$-band surface brightness evaluated within an outer galaxy ellipse corresponding to a 23.75 mag\,arcsec$^{-2}$ surface brightness level, the lowest-surface-brightness-level ellipse we find reliably determined). }}

Our completeness correction method results in multiplicative correction factors that vary as a function of absolute magnitude, color, and {\color{red}{$\mu_{23.75r}$ as illustrated in Fig.\ \ref{ccfield}. The median correction factor applied to a galaxy in our sample is $\sim$1\%, but correction factors can reach up to $\sim$68\%. Regions where correction factors are enhanced in this figure correspond to grid cells where genuinely large differences in the number counts of the comparison samples occur.}} Applying the derived completeness correction factors to ECO results in galaxy luminosity and color distributions that are in typically good agreement with RESOLVE-B sample properties. In addition to the {\color{red}{general redshift catalog}} incompleteness, we note that the presence of cluster fingers of God in the ECO sample can cause further incompleteness through the loss of members with extreme line-of-sight velocities, putting them outside our sample limits even though we include a large buffer region in our analysis to mitigate such losses. The presence of the Coma Cluster, in particular, leads to incompleteness for the highest group halo mass in the ECO sample (see upper leftmost panel in Fig.\ \ref{racz}), which cannot be appropriately corrected through consideration of the RESOLVE-B comparison sample as this incompleteness is caused by ECO boundaries and not incomplete redshift data.

\begin{figure*}
\epsscale{1.1}
\plottwo{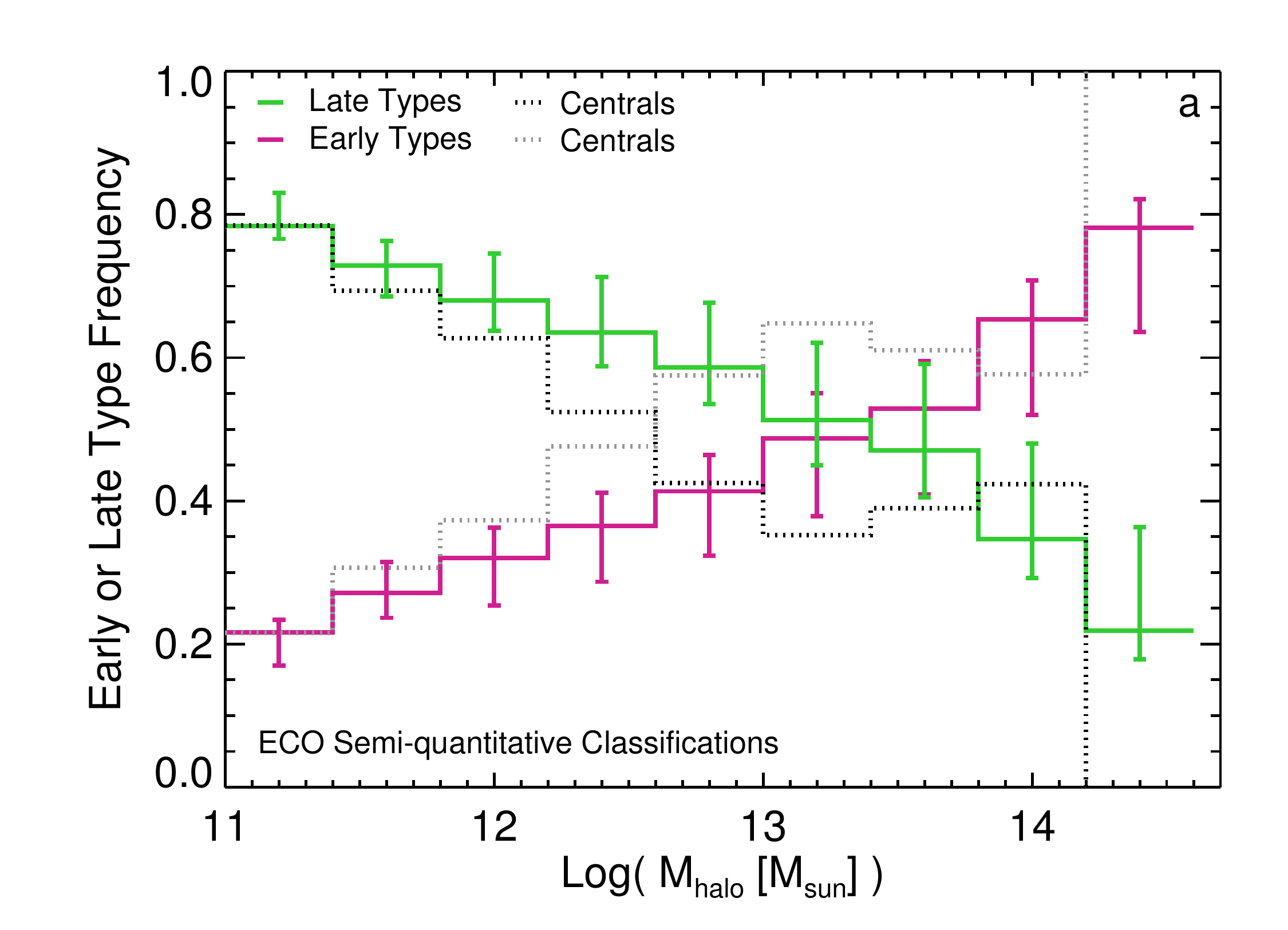}{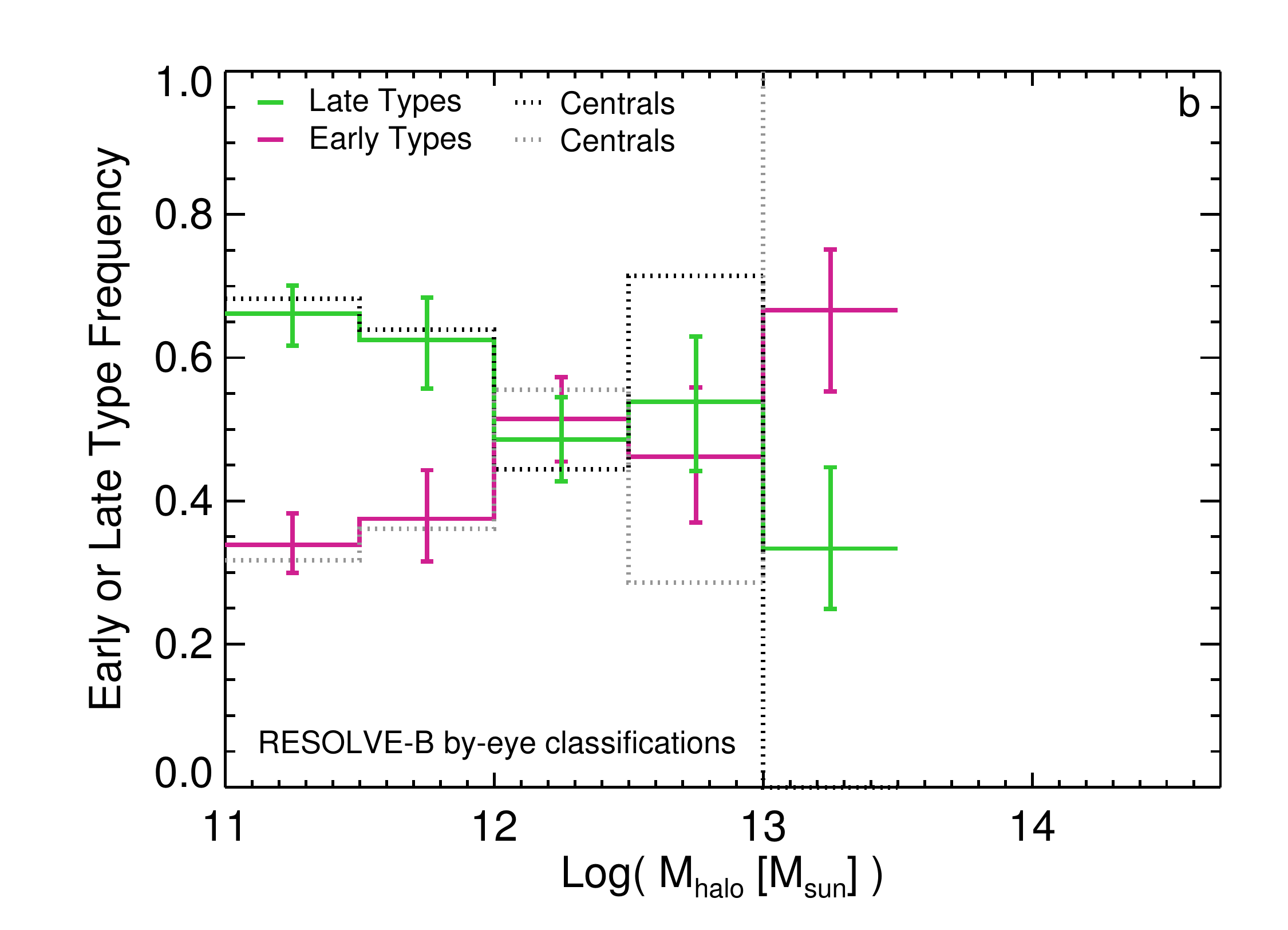}
\caption{Illustration of the traditional morphology-environment relation in the ECO and RESOLVE-B samples. Early and late type frequencies as a function of group halo mass in (a) ECO and (b) RESOLVE-B, with early and late type frequencies crossing over at $M_{halo} \sim 10^{13.5} M_{\odot}$ in the ECO sample. Greyscale dotted lines indicate the frequencies for central galaxies alone, which become poorly determined at high {\color{red}{group}} halo masses due to the small number of high halo mass groups (and thus centrals) present in the sample. In panel a, all frequencies are plotted at their ``expected'' value given the calibrated uncertainties in our semi-quantitative morphology classification method, described in \S \ref{morphclass}. The error bars in this panel are plotted as a combination of the calculated misclassification errors in each {\color{red}{group}} halo mass regime and the (binomial) counting statistics in each bin, while those in panel b reflect only the relevant counting statistics (assuming zero misclassification). Note that the early/late type frequencies for the two samples do not strictly agree in all bins, particularly near $M_{halo} \sim 10^{12} M_{\odot}$. Applying the quantitative classification approach to RESOLVE-B results in frequencies that more closely agree with the frequencies in ECO, although slight residual differences remain, perhaps resulting from group-to-group variations in typical properties.}
\label{MDA}
\end{figure*}

\begin{figure}
\epsscale{1.13}
\plotone{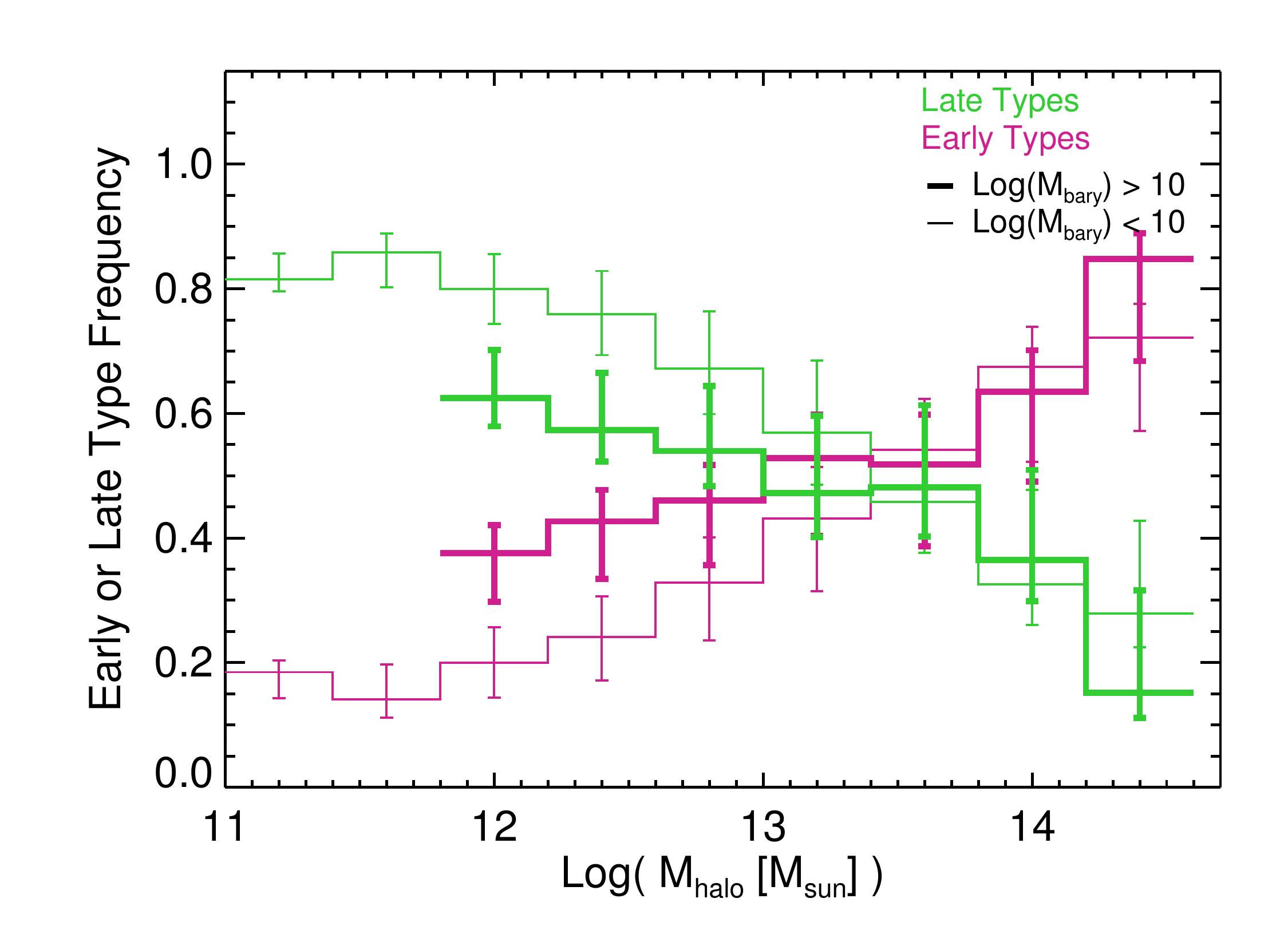}
\caption{Illustration of morphology-environment relations in the ECO sample for separate high-mass ($M_{bary} > 10^{10} M_{\odot}$) and low-mass ($M_{bary} < 10^{10} M_{\odot}$) galaxy subsamples, indicated by thick and thin lines, respectively. Since few high mass galaxies inhabit {\color{red}{group}} halos below $\sim10^{12} M_{\odot}$, we do not plot the frequencies for high mass galaxies below this point. While the trends are qualitatively similar over most of the group halo mass range, the relations have different amplitudes for the low and high galaxy mass samples.}
\label{MDAmass}
\end{figure}

To quantify correction factors for this boundary effect, we construct a comparison catalog including the same data sources as ECO but extending beyond the ECO redshift limits in both directions (1500 km/s to 12000 km/s), encompassing all apparent fingers of God extending outside of the ECO region. As reprocessed photometry is not available for this entire comparison catalog, we rely on SDSS catalog magnitudes and perform group finding {\color{red}{on the comparison catalog using catalog $r$-band magnitudes. We fix the linking lengths for this group finding to the values appropriate for the normal ECO region to avoid adverse effects from the fall-off in completeness beyond the ECO boundaries. For any galaxy in a group with N$>$10 that has been affected by proximity to the ECO volume edges, we calculate and apply an additional completeness correction factor based on the ratio of the number of galaxies in the full comparison catalog group to the number of galaxies within the ECO catalog boundaries, derived considering only galaxies with catalog magnitudes $M_{r} < -18.4$ (corresponding to the nominal SDSS $r$-band completeness limit at $cz =$ 12000 km/s). The boundary completeness correction factors are only necessary for two groups, with a maximum correction value of $\sim1.42$ for Coma and a correction value of $\sim1.14$ for the NGC4065 group (see Fig.\ \ref{racz}). We note that these correction values are of necessity approximate as they do not account for any possible population differences between the parts of a cluster interior and exterior to our sample volume. As they apply to only two group halos with $M_{halo} \gtrsim 10^{14} M_{\odot}$ in our sample, a regime in which it is already difficult for us to draw conclusions due to the small number of rich clusters sampled in a volume of ECO's size, we caution against over-interpreting results solely from this cluster regime. We will largely focus on more common moderate to low richness environments in this work. }}

As the loss of galaxies beyond ECO boundaries affects sample membership, this loss can also affect our derived group halo mass estimates. To correct for this effect, we use the same logic described above and apply a correction factor to each affected group halo mass based on the ratio of the total group luminosity derived in the comparison catalog to the total group luminosity derived in the ECO catalog. {\color{red}{The applied group halo mass correction for Coma is an increase of $\sim0.1$ dex but is negligible for the NGC4065 group}}.

\section{Results}
\label{res}

\begin{figure*}
\epsscale{1.15}
\plotone{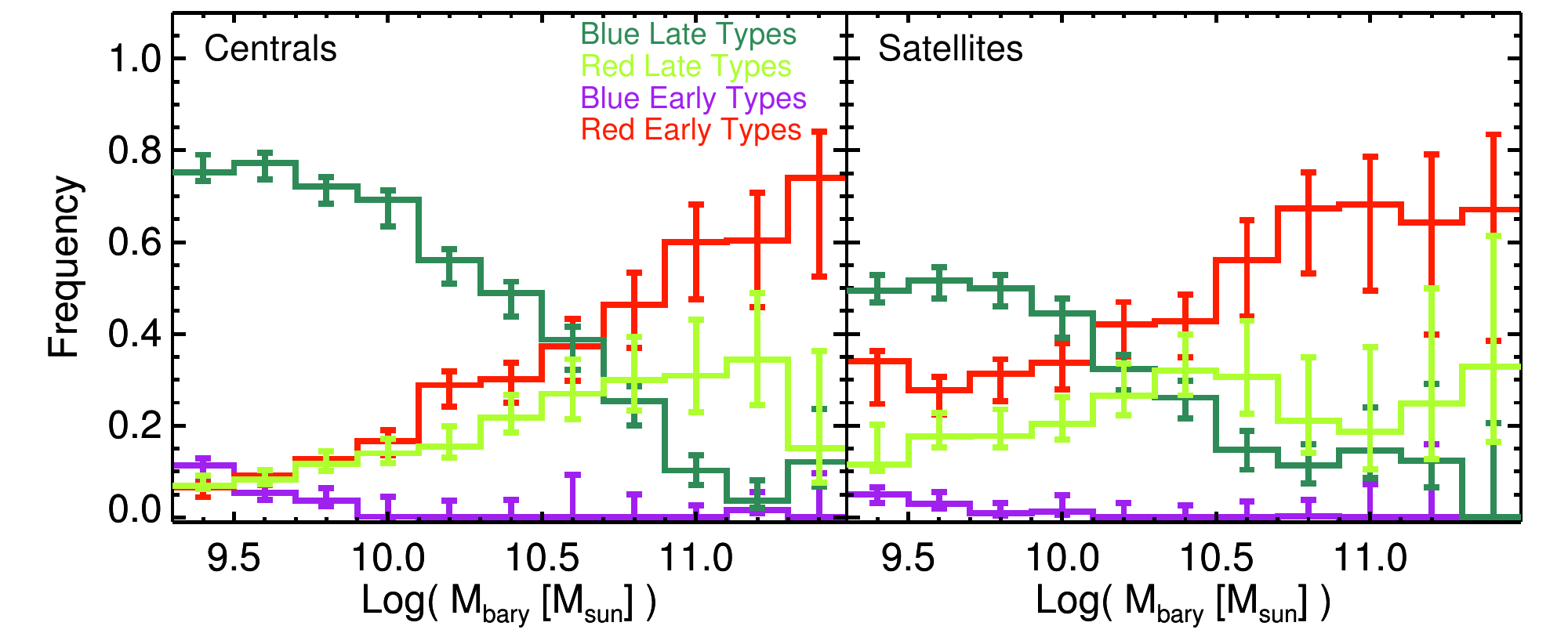}
\caption{Illustration of morphology-mass relations in the ECO sample for multiple color/morphology subclasses of ECO galaxies, where the left panel includes centrals only and the right panel includes satellites only.}
\label{MDmultpan_mass}
\end{figure*}

\begin{figure*}
\epsscale{1.15}
\plotone{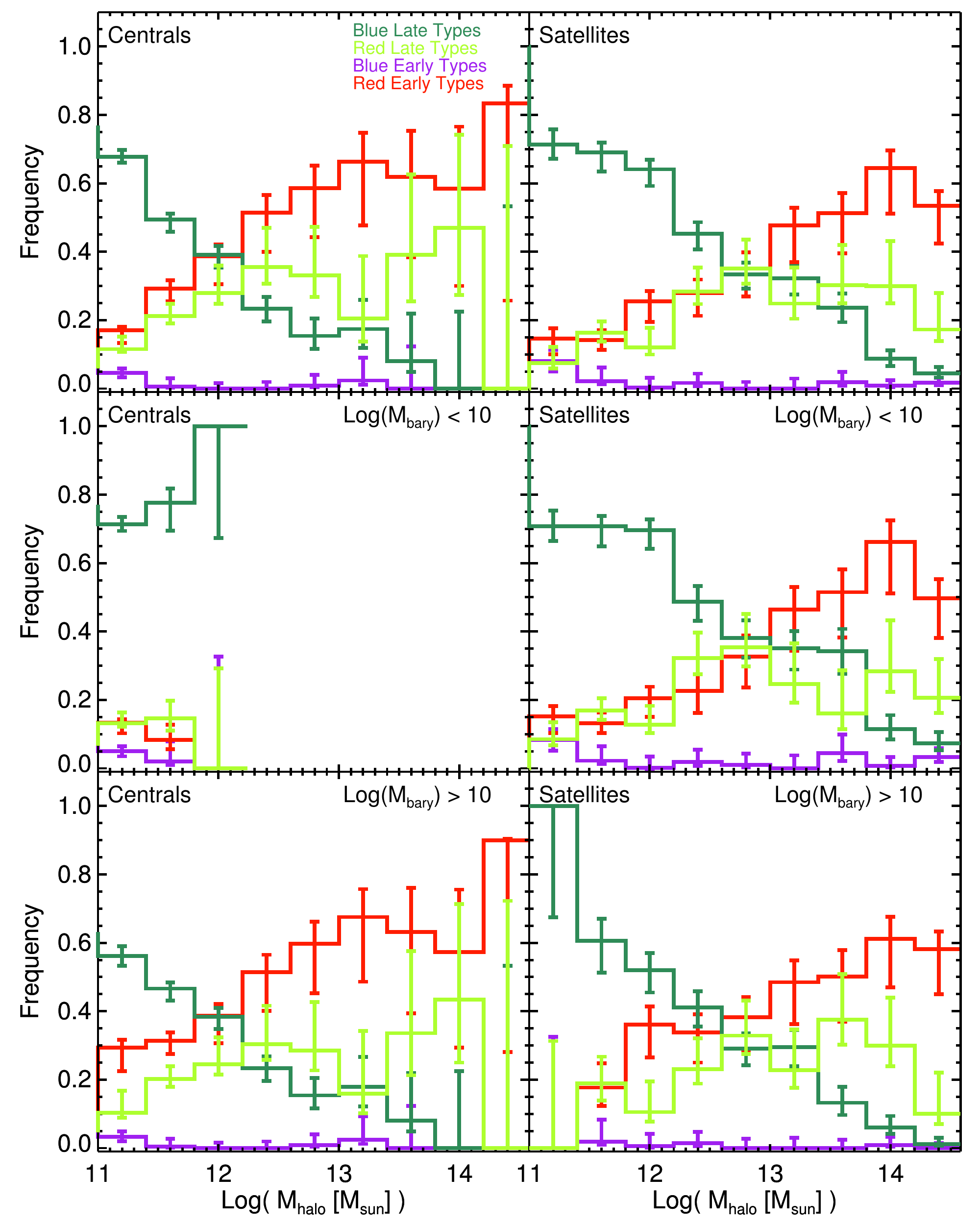}
\caption{Illustration of morphology-environment relations in the ECO sample for multiple subclasses of ECO galaxies. All plots in the left column include centrals only, while plots in the right column include satellites only. The bottom two panels show morphology-environment relations in the ECO sample for separate low-mass ($M_{bary} < 10^{10} M_{\odot}$) and high-mass ($M_{bary} > 10^{10} M_{\odot}$) galaxy subsamples.}
\label{MDmultpan}
\end{figure*}

  In this section, we explore trends in galaxy properties related to galaxy mass and environment with an eye towards their connections to galaxy disk (re)growth. {\color{red}{In all plots of our results we include only objects that meet our final group membership and $M_{bary} > 10^{9.3} M_{\odot}$ selection as detailed in \S \ref{samp}.}} We use group halo mass as our primary indicator of environmental richness. {\color{red}{As mentioned previously in \S \ref{gfsec}, we also consider isolated galaxies as members of groups with N=1. Although this description may at first seem counterintuitive, it does provide a way to quantify group environment as a continuous variable through group halo mass, which is useful as a basis for comparison with theory. Since the $r$-band luminosity of a galaxy provides an excellent proxy for its total \emph{baryonic} rather than stellar mass (as discussed further in \citealp{K13}), it is the case that where group N=1 our estimated group halo mass is directly related to the lone galaxy's baryonic mass. We can nonetheless identify group halo environment rather than galaxy mass as the primary driver of a trend if it is sensitive to group central versus satellite status at fixed galaxy mass.}} 

The distribution of various classes of ECO catalog galaxies in group halo mass versus galaxy baryonic (stellar plus atomic gas) mass parameter space can be seen in Fig.\ \ref{mhalomstarscat}. Results derived from the density field are qualitatively similar to results derived from group halo masses, except as discussed in this section.

\subsection{Traditional Morphology-Environment Relation - P(M$|$E)}

  As seen in Fig.\ \ref{MDA}a for the traditional formulation of the morphology-environment relation (frequency of a particular morphology as a function of environmental richness), the ECO sample displays the expected increase in early-type and decrease in late-type frequencies as a function of increasing environmental richness, here represented by increasing group halo mass. Near group halo mass $\sim$~$10^{13.5} M_{\odot}$, we observe a crossover point where early and late type frequencies become approximately equal. The frequencies in the more complete RESOLVE-B sample, where morphological classifications are entirely based on by-eye judgments, do not strictly agree with the frequencies in ECO, particularly around $M_{halo} \sim 10^{12} M_{\odot}$ (see Fig.\ \ref{MDA}b). If we apply quantitative classification methods to RESOLVE-B, we find frequencies that more closely but not completely agree with the frequencies in ECO. Variations in morphological mixtures between groups at fixed halo mass could also plausibly contribute to differing {\color{red}{average}} morphological mixes in these two samples. If we divide central and satellite galaxies, we find that central and satellite morphology-environment trends are similar to each other in both samples, where the satellite galaxy trend (not shown) closely follows the combined trend in Fig.\ \ref{MDA}a/b. If we examine the traditional morphology-environment relation for low and high baryonic mass (divided at $M_{bary} = 10^{10} M_{\odot}$) galaxies separately, we qualitatively recover the expected early-/late-type frequency trends as a function of {\color{red}{group halo mass, with the high baryonic mass relations shallower than the low baryonic mass relations and offset towards higher early-type frequency. The early/late-type frequencies cross over at group halo masses $\gtrsim10^{13} M_{\odot}$, and both trends steepen in this regime (see Fig.\ \ref{MDAmass}). }}

  {\color{red}{Dissecting the morphology-environment and morphology-mass relations further by including galaxy color, we can see from Fig.\ \ref{MDmultpan_mass} that the offset between low and high baryonic mass morphology-environment relations must be partially driven by the significant changes in blue and red galaxy frequencies as a function of galaxy mass, specifically as blue late types are significantly more common at low galaxy mass and red early and late types are more common at high galaxy mass. In Figs.\ \ref{MDmultpan_mass} and \ref{MDmultpan}, we also see that centrals and satellites within each galaxy color/morphology class follow qualitatively similar trends as a function of galaxy mass and environment. However several notable offsets in frequency occur, for example in Fig.\ref{MDmultpan_mass}, blue late types are less common among satellites than centrals at low baryonic mass, whereas red early types are more common among satellites than centrals at low baryonic mass. Interestingly, in the top panels of Fig. \ref{MDmultpan}, blue late types are \emph{more} common among satellites than centrals at low \emph{group halo} mass, illustrating that galaxy demographic trends in the low galaxy mass and low group halo mass regimes are not necessarily equivalent. There is also a hint of a similar reversal for red early types (i.e., red early types are marginally more common among centrals than satellites at low group halo mass). As can be seen in the bottom two panel sets in Fig.\ \ref{MDmultpan}, the galaxy color/morphology class frequency trends as a function of group halo mass, where they can be quantified, are overally similar between low and high galaxy mass subsamples.}}

\begin{figure*}
\epsscale{1.1}
\plottwo{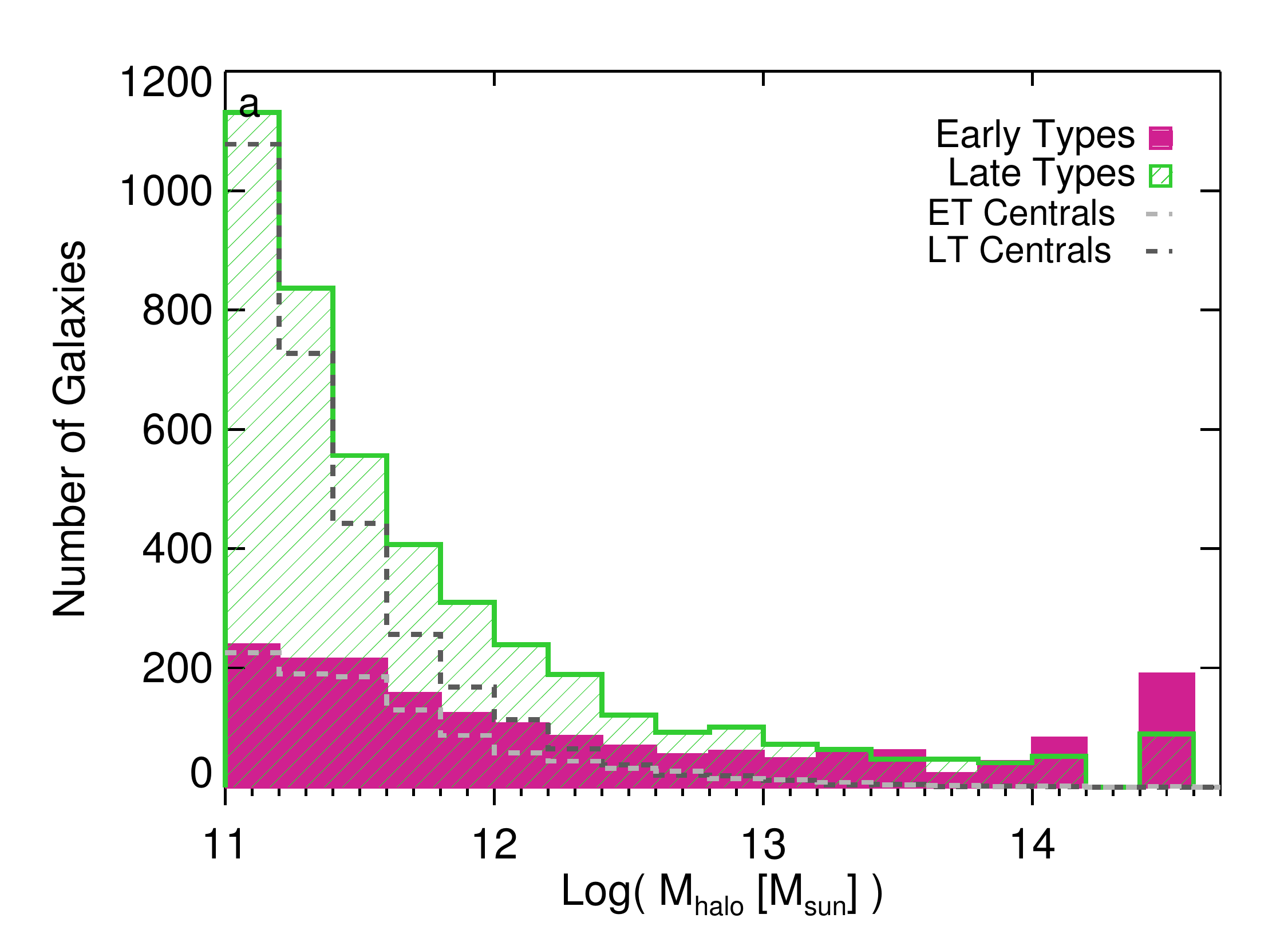}{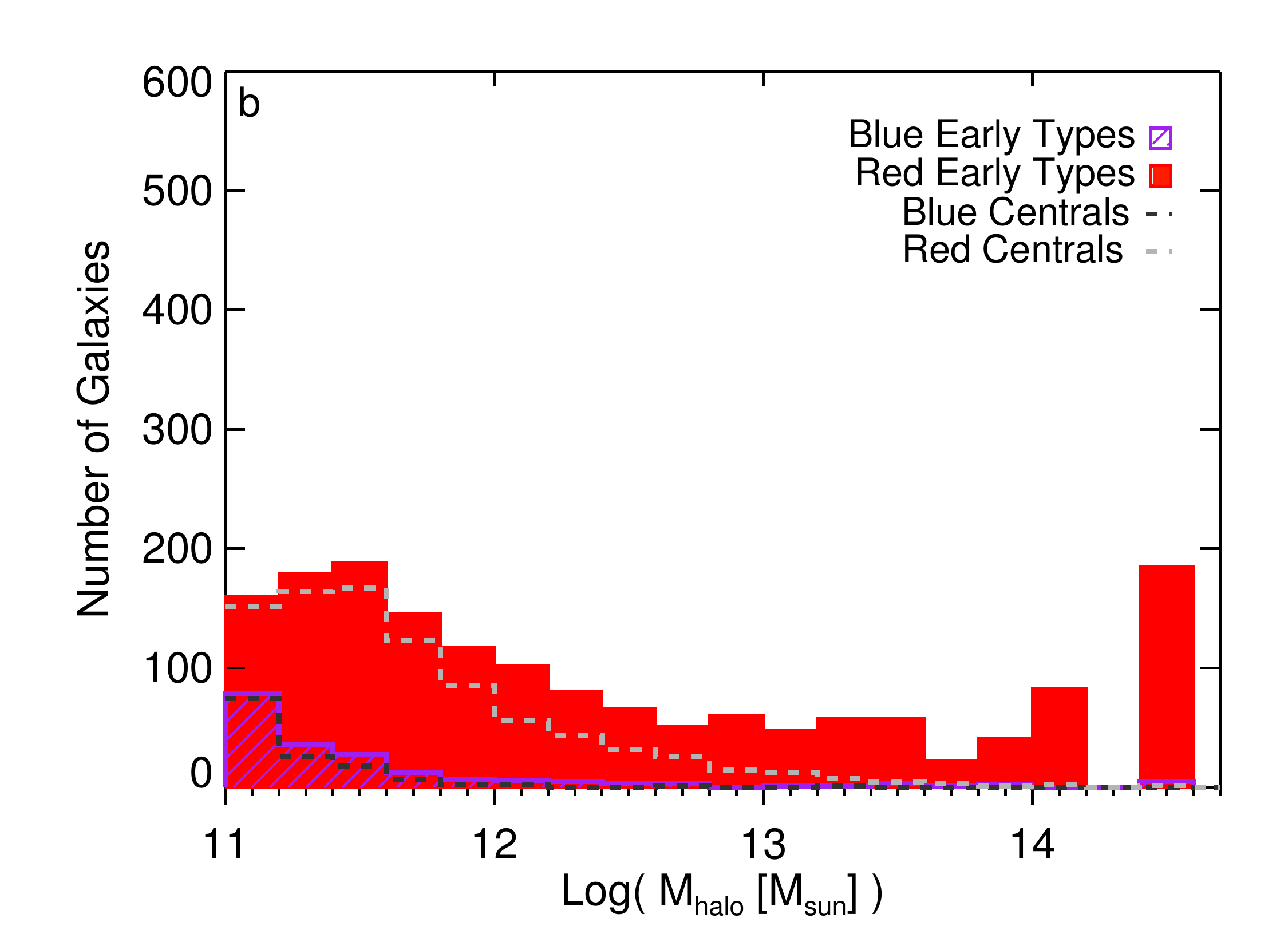}
\caption{{\color{red}{Group}} halo mass distribution for early and late type galaxies in the ECO sample, with corresponding grey dashed lines indicating the distribution for group central galaxies of each type. Panel \emph{a} shows all early and all late types together, and panel \emph{b} breaks down the early types further by red and blue sequence membership. Approximately 50\% of ECO early type galaxies occupy $M_{halo} < 10^{12} M_{\odot}$ environments, approximately 20\% of which are blue early types.}
\label{MDB}
\end{figure*}

\begin{figure}
\epsscale{1.13}
\plotone{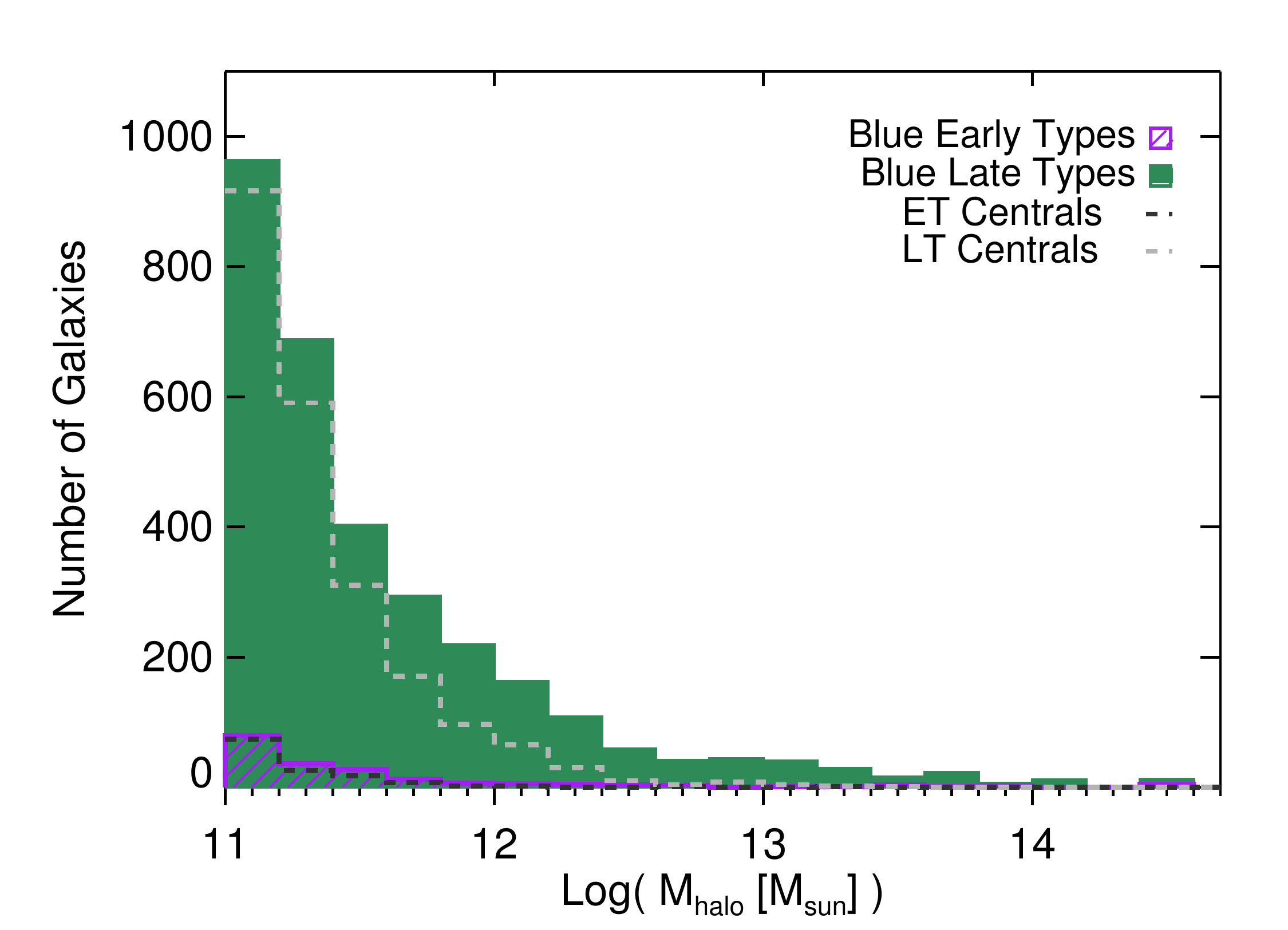}
\caption{{\color{red}{Group}} halo mass distribution for blue early-type and blue late-type galaxies in the ECO sample, with corresponding grey dashed lines indicating the distribution for group central galaxies of each type.}
\label{MDB2}
\end{figure}

{\color{red}{
\subsubsection{Focus on Blue-sequence Early Type Galaxies}

As illustrated in Fig.\ \ref{MDB}a, the majority of late-type galaxies occupy $M_{halo} < 10^{12} M_{\odot}$ environments, which is not surprising given the form of the traditional morphology-environment relation. More interestingly the same figure shows that $\sim$50\% of early-type galaxies occupy $M_{halo} < 10^{12} M_{\odot}$ environments.}} As seen in Fig.\ \ref{MDB}b, many early types in the lowest density environments are \emph{blue-sequence} early types. These blue early types become most common below $M_{halo} \sim 10^{11.5} M_{\odot}$. Blue early types with baryonic masses large enough to meet our mass limit are primarily central galaxies in this low group halo mass regime and primarily satellite galaxies in richer environments (see Fig.\ \ref{mhalomstarscat}). However, since satellite blue early types with masses below our survey limit could also populate the low mass halos, we cannot yet quantify the balance between blue early type centrals and satellites in these environments. Fig.\ \ref{MDB2} shows that blue early-type galaxies appear to have similar environment distributions to blue late-type galaxies overall, {\color{red}{but blue early types display a slight preference towards lower group halo mass environments. As calculated with the Kolmogorov-Smirnov (K-S) test, the group halo mass distributions of blue early and blue late types in Fig.\ \ref{MDB2} have $\sim1\%$ probability of being drawn from the same parent distribution ($P_{same} \sim 0.01$).}}

\begin{figure}
\epsscale{1.16}
\plotone{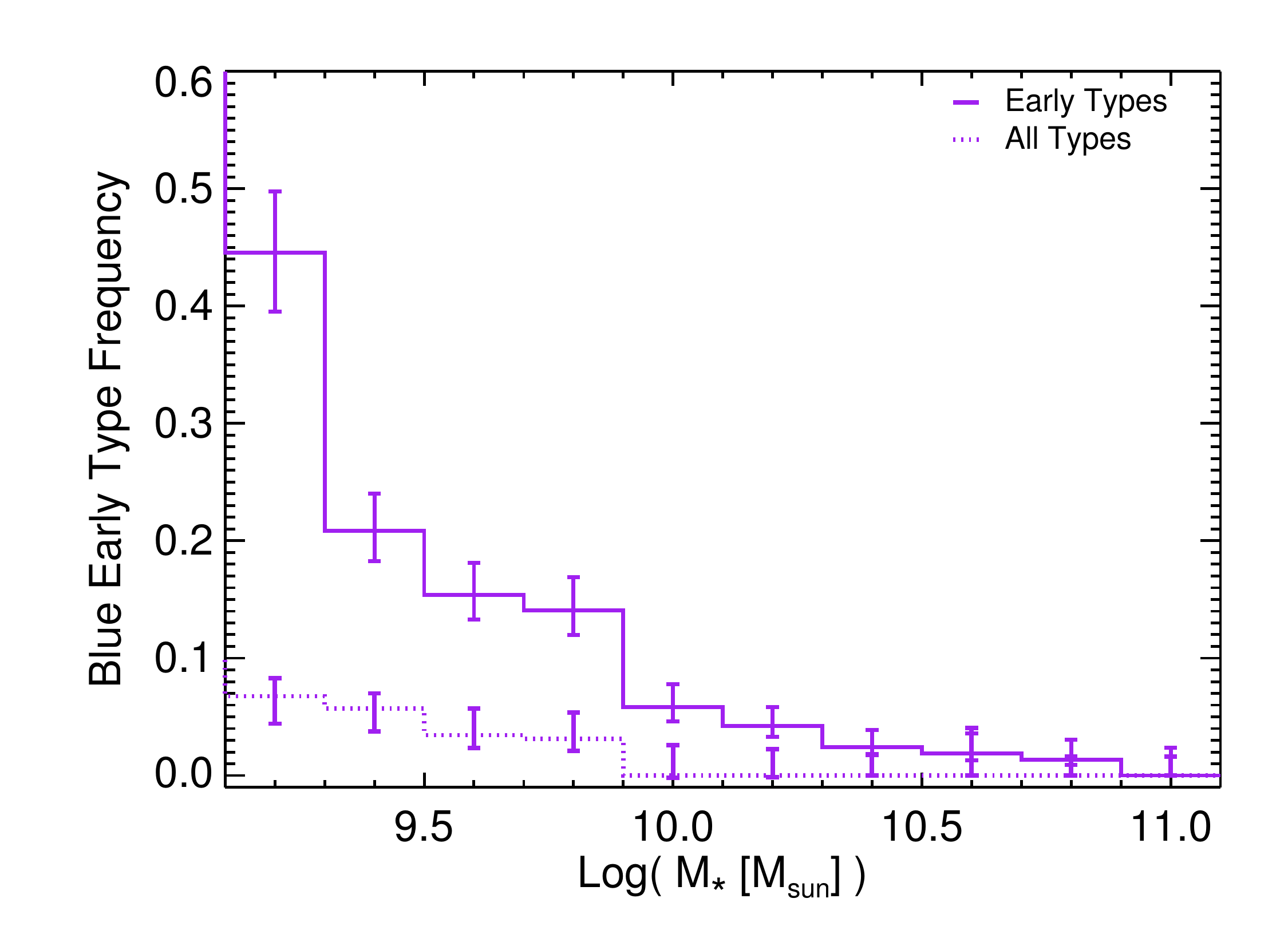}
\caption{Variation in blue early type galaxy frequency as a function of stellar mass in the ECO sample. The solid line indicates the frequency of blue early types as a fraction of early types only, while the dotted line indicates their frequency as a fraction of all galaxy types. Frequencies are plotted at their expected values given the calibrated uncertainties in our semi-quantitative morphology classification method, described in \S \ref{morphclass}. Error bars shown are a combination of the estimated misclassification errors and the (binomial) counting statistics in each bin.}
\label{betm}
\end{figure}

We find that blue-sequence early-type galaxies are not only most common in low {\color{red}{group}} halo mass environments but at low {\color{red}{galaxy}} masses as well, both of which are regimes where extreme gas richness is typical. As was previously found by KGB, we observe that the blue early-type galaxies in ECO become more common with \emph{decreasing} stellar {\color{red}{or baryonic mass (Figs.\ \ref{betm} and \ref{MDmultpan}).}} These galaxies only emerge in ECO with measurable frequency around the galaxy bimodality mass ($M_{b} \sim 10^{10.5} M_{\odot}$) and increase in frequency significantly below $M_{*} \sim 10^{10} M_{\odot}$. This behavior is similar to that observed in KGB, where the blue early-type population increases sharply below $M_{*} \sim 10^{9.7} M_{\odot}$, the ``gas-richness threshold'' mass. Our observed frequency transition is somewhat less sharp than that observed by KGB, but it is plausible that any sharp transitions could be washed out by the error rates inherent in our semi-quantitative classification method. Our absolute frequencies are similar to those reported by KGB, although not in strict agreement within our error bars. If similarly large error bars on the KGB frequencies are assumed, then the two trends would agree overall. \citet{Thomas10} have also observed an increasing frequency of ``rejuvenated'' early-type galaxies with decreasing galaxy mass (and environmental density), reaching a maximum of $\sim$45\% of the early-types, which is similar to our observed maximum frequency. This observed low-mass preference implies that mass-dependent mechanisms are closely tied to the rise of the blue-sequence early-type population. We note that the typically low stellar mass of blue early-type galaxies appears linked to the typically low group halo mass environments they inhabit. At $M_{*} \sim 10^{10} M_{\odot}$, the typical baryonic mass for a galaxy is {\color{red}{$\sim10^{10.1}$ to $10^{10.2} M_{\odot}$}}, which from Fig.\ \ref{mhalomstarscat} corresponds to a typical {\color{red}{group}} halo mass of $\sim10^{11.5} M_{\odot}$ for centrals.

\subsection{Alternative Morphology-Environment Relation - P(E$|$M)}

\begin{figure*}
\plotone{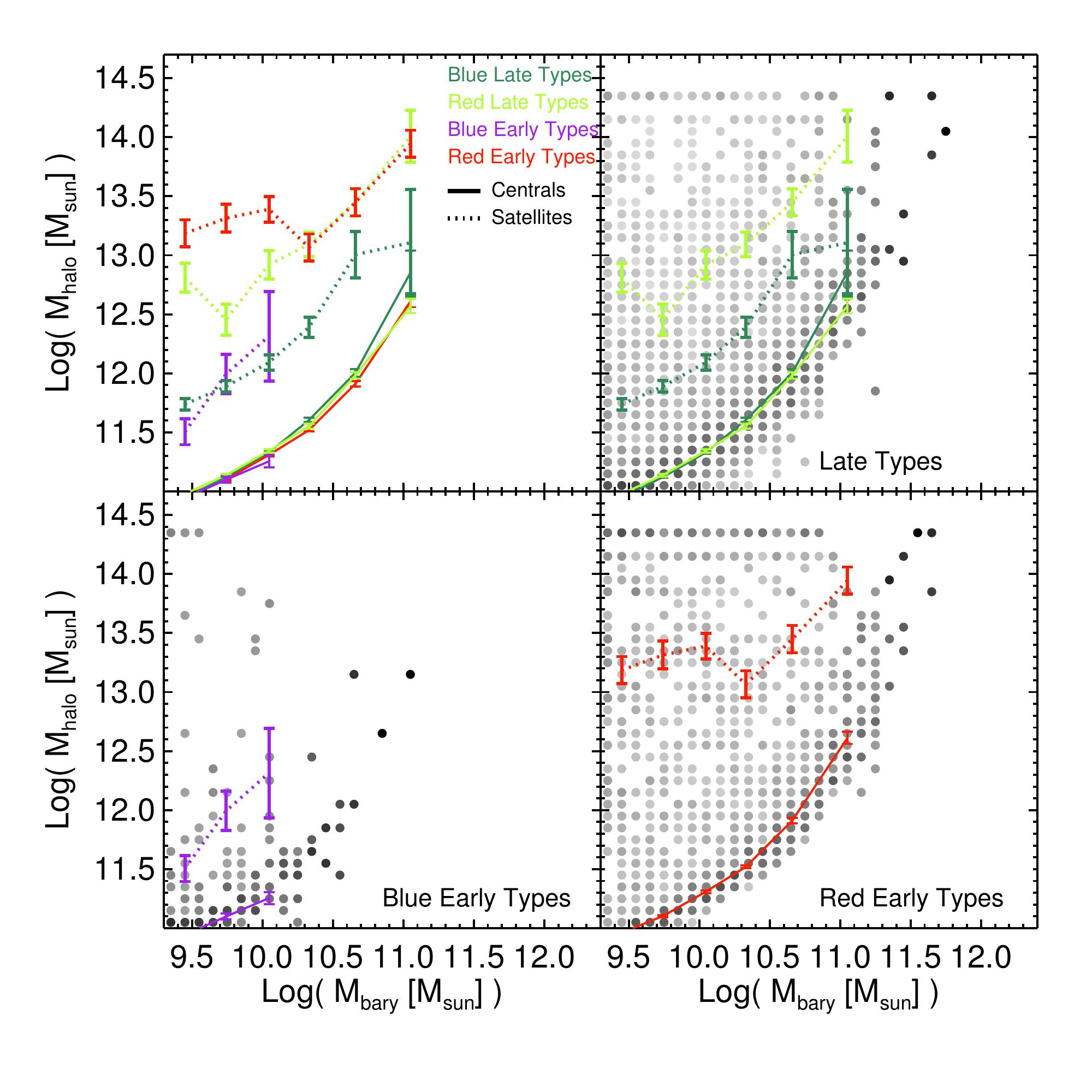}
\caption{Characteristic distribution of {\color{red}{group}} halo mass as a function of baryonic mass for different galaxy types. Solid lines indicate the running median group halo masses for centrals of each galaxy type, while dashed lines indicate the medians for satellites of each galaxy type. Error bars for each median point are estimated from the dispersion in properties in each bin. The background greyscale levels indicate the probability of inhabiting a particular {\color{red}{group}} halo mass at a given baryonic mass, as the histogram densities used to set the greyscale have been normalized to one in each baryonic mass bin (darkest points imply the highest probabilities).}
\label{mhalombarydsym}
\end{figure*}

{\color{red}{An alternative formulation of the morphology-environment relation, expressing the probability for a galaxy with a given morphology to inhabit a particular environment, is illuminating for understanding the ``typical'' environments of various galaxy classes as a function of galaxy mass as illustrated in Fig. \ref{mhalombarydsym}. As seen in this figure the typical environments of blue early types and blue late types are similar at constant baryonic mass, for both centrals and satellites. This figure also illustrates the general lack of a P(E$|$M) morphology-environment relationship for central galaxies at fixed mass, which also holds if we consider all early types and all late types together and is essentially a manifestation of the known central galaxy mass to group halo mass relation (e.g., \citealp{Yang2009p3}; \citealp{Moster2010}). 

The traditional P(M$|$E) morphology-environment relation obscures this fact by mixing together morphological classes with differing baryonic mass distributions within each environment regime. However, for low baryonic mass ($M_{bary} < 10^{10} M_{\odot}$) galaxies taken alone even the traditional morphology-environment relation is approximately flat until group halo masses above $\sim$10$^{12.5} M_{\odot}$ (see Fig.\ \ref{MDAmass}), about 1 dex into the regime where these low mass galaxies have become satellites. Accordingly, in the P(E$|$M) formulation, a morphology-environment relation reemerges for low baryonic mass satellites: in general early-type satellites typically occupy higher group halo mass environments than late-type satellites at fixed baryonic mass, although specifically this relation is driven by \emph{red} early types. For satellite galaxies, the strongest difference in typical group halo mass is between red galaxies and blue galaxies, where both red early and late types occupy higher group halo mass environments than either blue early or late types.}}

\begin{figure*}
\plotone{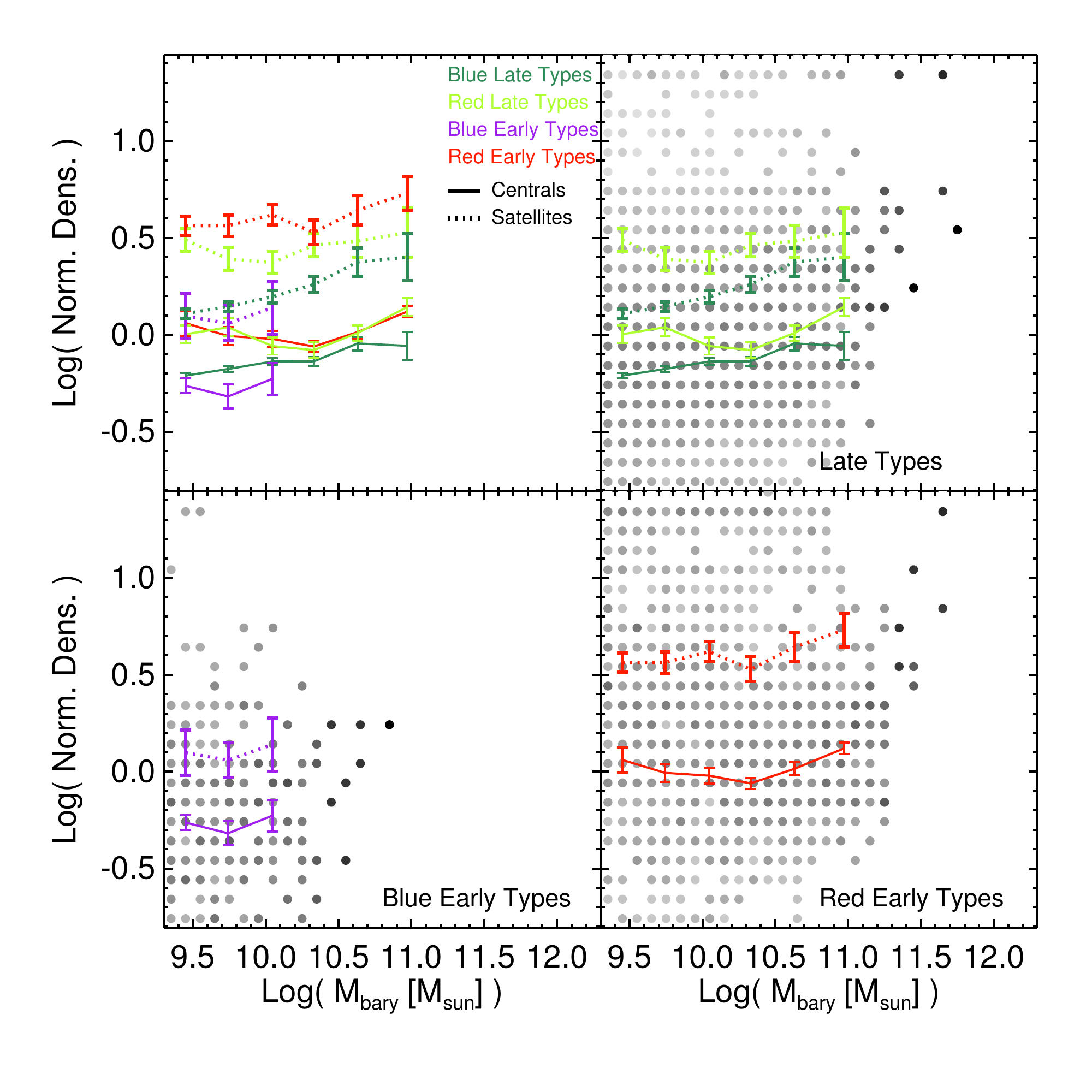}
\caption{Characteristic distribution of environmental density ($\sim$1.43 Mpc smoothing kernel) as a function of baryonic mass for different galaxy types, with symbols and lines analogous to Fig.\ \ref{mhalombarydsym}.}
\label{densmstar}
\end{figure*}

Even though low mass blue early and blue late types occupy environments that are typically similar {\color{red}{(and are classified as centrals with the same $\sim$80\% frequency for $M_{bary} < 10^{10} M_{\odot}$)}}, their full environment probability distributions at a given mass may not necessarily be the same. In this case, we find K-S test $P_{same} \sim 0.22$ for the group halo mass distributions of $M_{bary} \lesssim 10^{10} M_{\odot}$ blue early and blue late types, where the blue late type environment distribution appears to be broader, {\color{red}{suggesting greater environmental diversity at fixed mass (although we note that the blue late type distribution is better sampled).}} Considering environmental density values (smoothed on $\sim$1.43 Mpc scales), we find that the typical densities around $M_{bary} \lesssim 10^{10} M_{\odot}$ blue early-type centrals are lower than those of late types (see Fig.\ \ref{densmstar}a). If number density is more sensitive to major mergers than group halo mass, then this tendency towards somewhat lower environmental densities could be consistent with blue early-type centrals existing as post-merger objects with their number of neighbors reduced by merging. As with group halo masses, blue early-type and blue late-type satellites have similar typical densities in the  $M_{bary} \lesssim 10^{10} M_{\odot}$ regime, but the overall blue early and blue late type density distributions in this mass regime are {\color{red}{more plausibly different}}, with a $P_{same} \sim 0.06$ as quantified with the K-S test.

\subsection{Extreme Gas Richness and Environment}

\begin{figure}
\epsscale{1.1}
\plotone{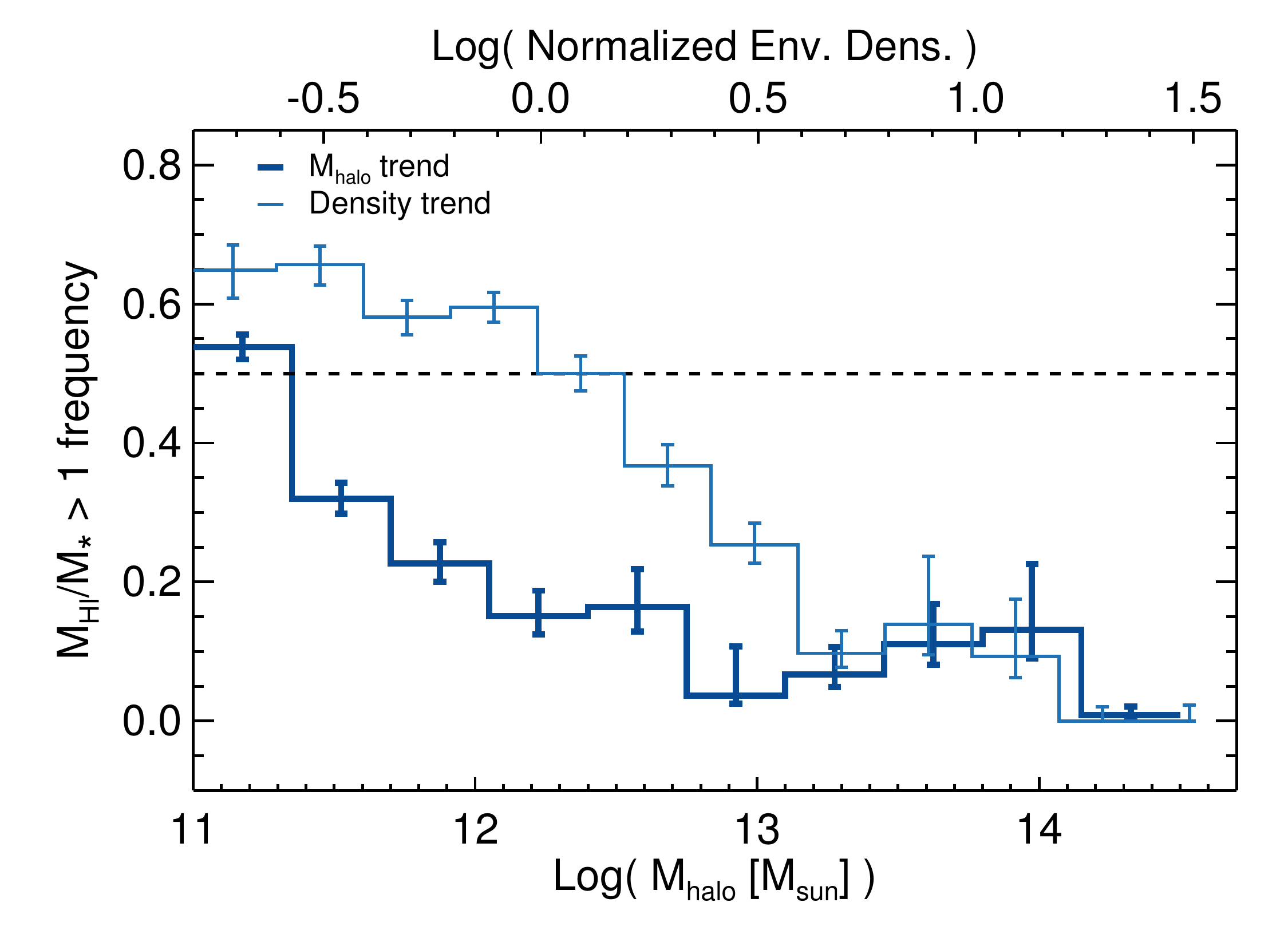}
\caption{The environment-dependent frequency of HI gas-to-stellar mass ratios greater than one for ECO+A galaxies. The vast majority of ALFALFA $\alpha.40$ derived HI limit values for ECO galaxies are sufficiently strong to place these galaxies in either the $M_{HI}/M_{*} > 1$ or $M_{HI}/M_{*} < 1$ category, but for any limit values found in the $M_{HI}/M_{*} > 1$ category, we replace these values with photometric gas fraction estimates as described in \S \ref{HIest}. If we leave the limit values in place and use no photometric gas fraction estimates, we still find trend lines that are within the error bars of the trends shown here. Confused sources in the ECO+A sample have been omitted. The thick dark blue line illustrates the sharp uptick of the gas-dominated galaxy frequency at low group halo mass (lower x axis), and the thin light blue line illustrates the smoother increase in gas-dominated galaxy frequency as a function of environmental density (upper x axis). The horizontal dashed line indicates equal $M_{HI}/M_{*} > 1$ and $M_{HI}/M_{*} < 1$ galaxy frequency.}
\label{HIfrac}
\end{figure}

\begin{figure}
\epsscale{1.1}
\plotone{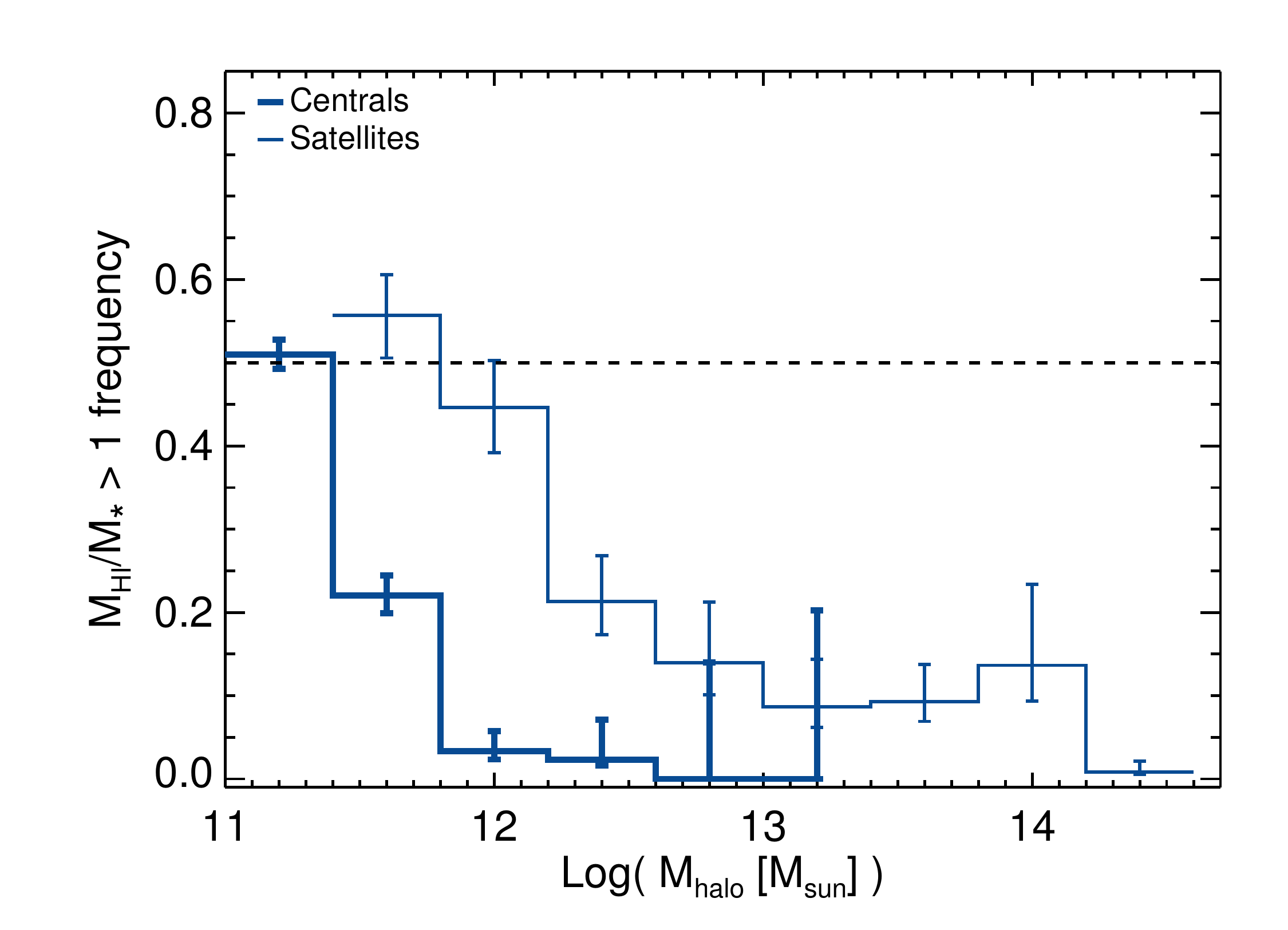}
\caption{The frequency of HI gas-to-stellar mass ratios greater than one for ECO+A galaxies alone, plotted as a function of group halo mass for central (thick lines) and satellite (thin lines) galaxies separately. The horizontal dashed line indicates equal $M_{HI}/M_{*} > 1$ and $M_{HI}/M_{*} < 1$ galaxy frequency. Confused sources in the ECO+A sample have been omitted. Since ECO contains relatively few satellite galaxies in $M_{halo} \lesssim 10^{11.5} M_{\odot}$ environments, we refrain from plotting satellite frequencies in this regime. Since ECO contains relatively few centrals above $M_{halo} \gtrsim 10^{13.5} M_{\odot}$, we likewise refrain from plotting central frequencies in this regime. Gas-dominated galaxy fraction increases significantly for low group halo mass central galaxies, particularly at $M_{halo} \lesssim 10^{11.5} M_{\odot}$.}
\label{HIfraccs}
\end{figure}

\begin{figure*}
\epsscale{1.}
\plotone{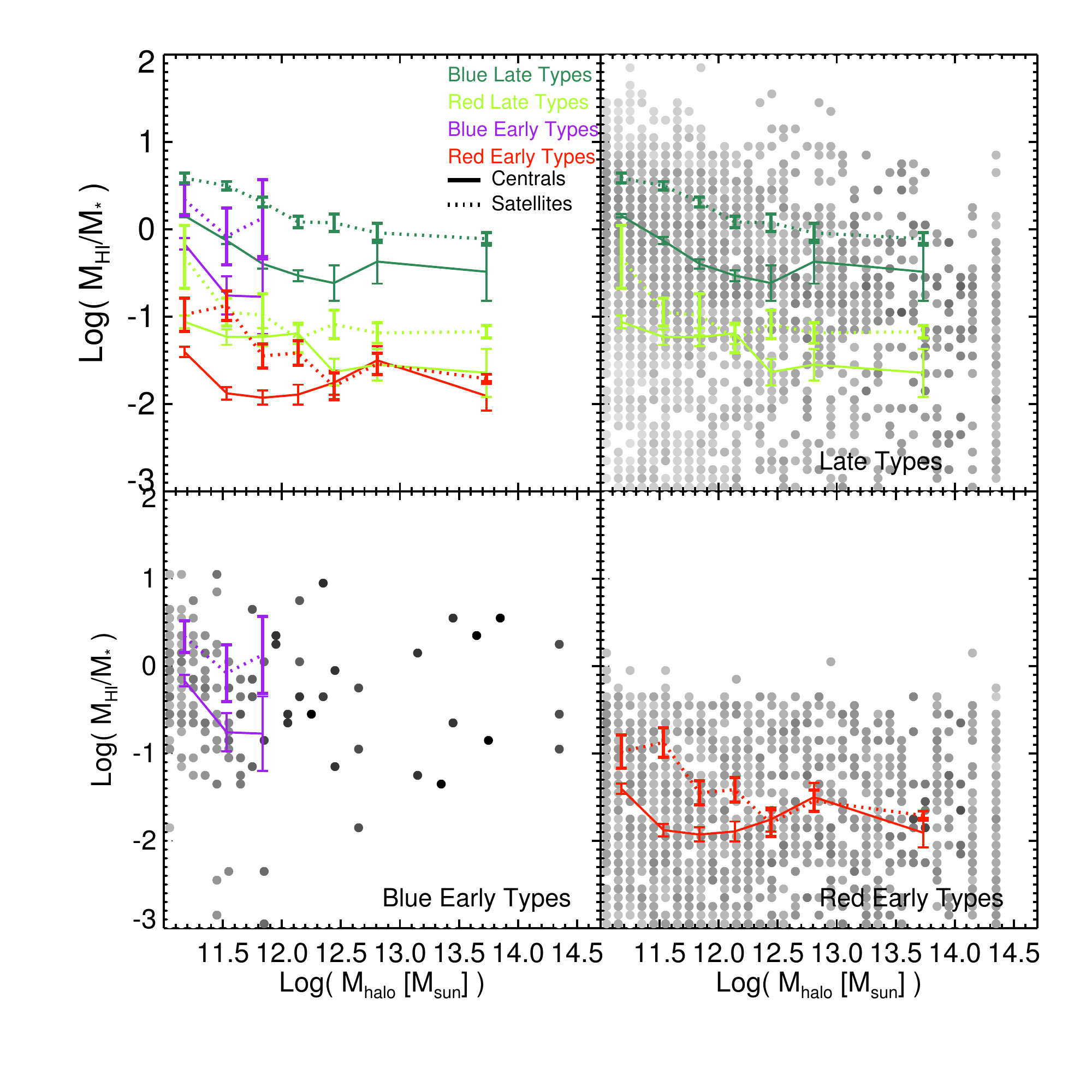}
\caption{Characteristic distribution of HI gas-to-stellar mass ratio as a function of group halo mass for different galaxy types. HI gas masses are derived as described in \S \ref{HIest}. Symbols and lines are analogous to Fig.\ \ref{mhalombarydsym}.}
\label{HImhalodsym}
\end{figure*}

\begin{figure}
\epsscale{1.15}
\plotone{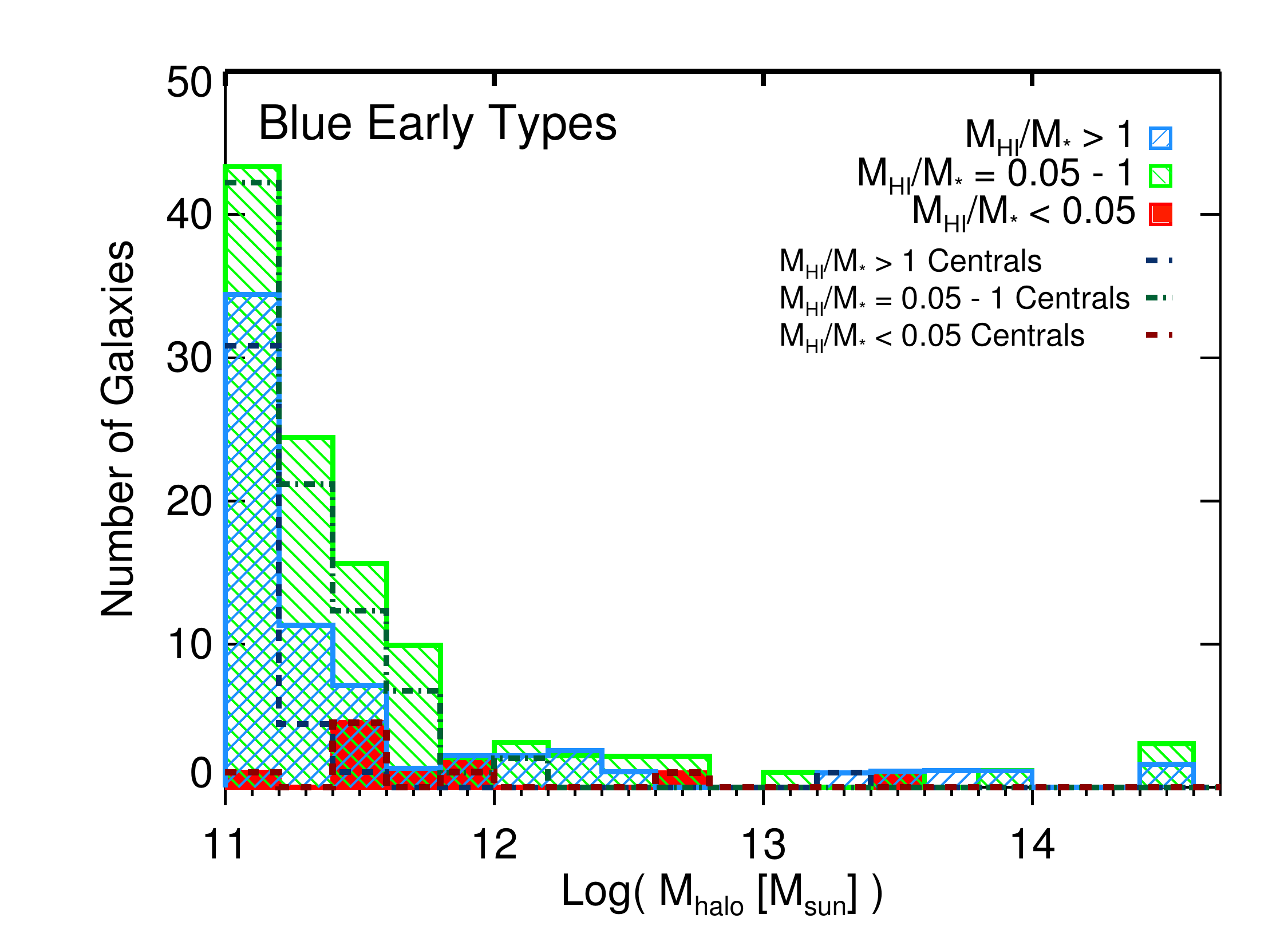}
\caption{{\color{red}{Group}} halo mass distribution for blue early type galaxies in the ECO sample with different levels of HI gas. Corresponding greyscale dashed lines indicate the distribution for group central galaxies of each type. HI masses are derived as described in \S \ref{HIest}. Gas-dominated blue early types and blue early types with more moderate gas-to-stellar mass ratios primarily inhabit environments with group halo masses below $\sim 10^{12} M_{\odot}$, with $\sim$53\% of gas-dominated blue early types in $M_{halo} \lesssim 10^{11.5} M_{\odot}$ environments.}
\label{betgashist}
\end{figure}

\begin{figure*}
\epsscale{1.}
\plotone{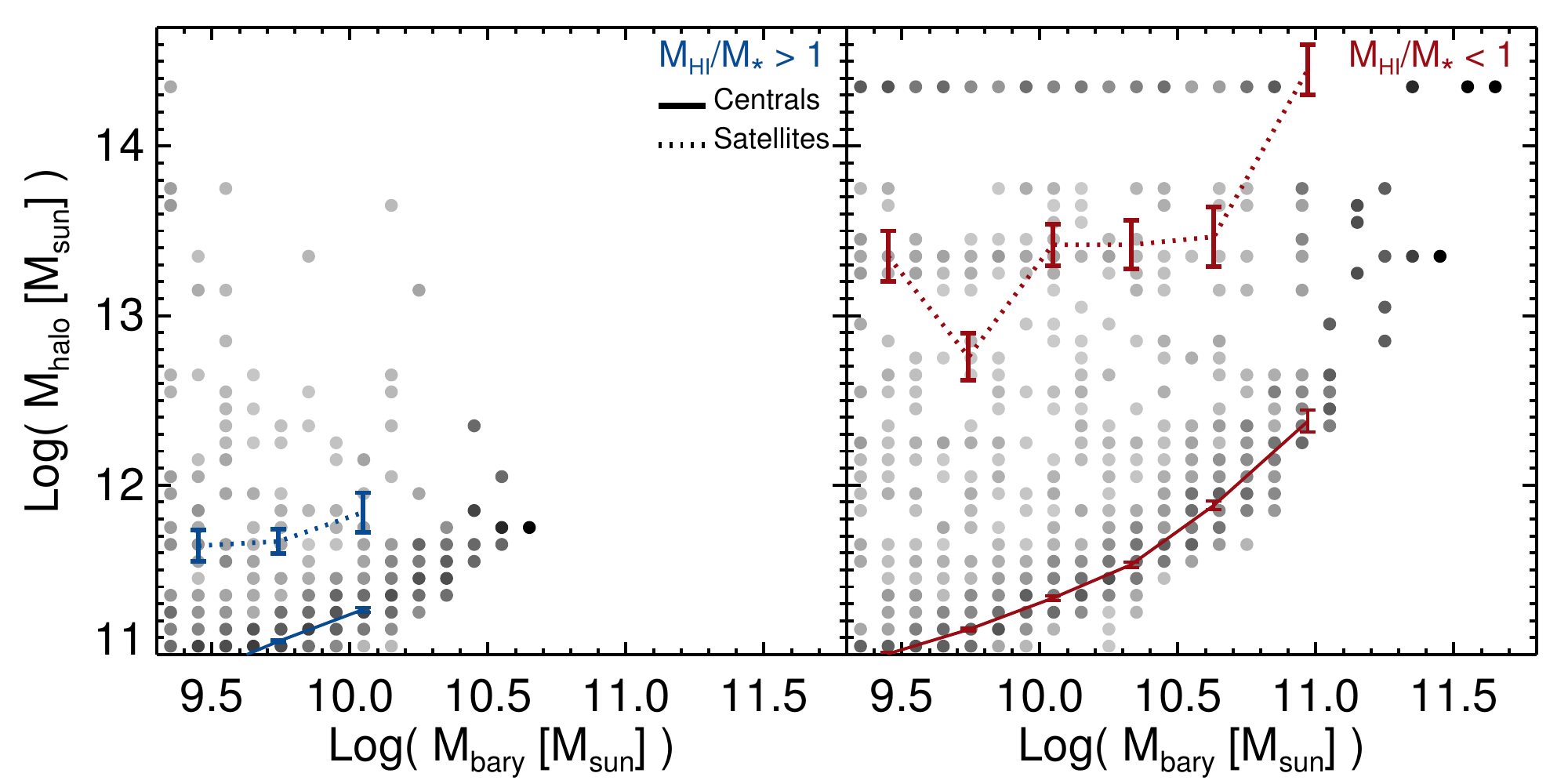}
\caption{Illustration of the typical galaxy and group halo mass distributions for ECO galaxies with HI gas-to-stellar mass ratios greater and less than one. HI gas masses are derived as described in \S \ref{HIest}. At fixed baryonic mass, high gas fraction satellites tend to occupy lower mass group halos than do low gas fraction satellites, implying group halo mass affects satellite gas content more strongly than galaxy mass.}
\label{HImhalombaryfrac}
\end{figure*}

  Considering ECO+A galaxies of all types together, Fig.\ \ref{HIfrac} shows that the fraction of gas-dominated galaxies (i.e., those with $M_{HI}/M_{*} > 1$) is a strong function of environment in general. The shape of the trend differs as a function of group halo mass compared to smoothed density field values: the halo mass relation displays a relatively smooth rise in gas-dominated galaxy frequency with a steep rise below $\sim10^{11.4} M_{\odot}$, whereas the density relation shows an overall smooth rise in gas-dominated galaxies toward lower densities (compare two trends in Fig.\ \ref{HIfrac}). The sharp increase in the frequency of extreme gas richness in our sample below $M_{halo} \sim 10^{11.4} M_{\odot} $ marks this as a regime where fractionally large gas reservoirs become a common feature of galaxies. Furthermore, we see that central galaxies are primarily responsible for the sharp increase in gas-dominated galaxy frequency with {\color{red}{group}} halo mass, within the baryonic mass limits of our sample (see Fig.\ \ref{HIfraccs}). 

If we consider the typical $M_{HI}/M_{*}$ values as a function of environment for individual galaxy types in the full ECO sample, we find that in low {\color{red}{group}} halo mass environments blue early types display $M_{HI}/M_{*}$ values that are comparable or somewhat lower than those of blue late types (see Fig.\ \ref{HImhalodsym}). We also observe that gas-dominated blue early types, with HI gas-to-stellar mass ratios $>1$, are found almost entirely in low {\color{red}{group}} halo mass environments (see Fig.\ \ref{betgashist}), becoming most common below a group halo mass of $\sim10^{11.5} M_{\odot}$, where they are primarily group centrals (although again, any satellites would likely fall below our mass limit).

From Fig.\ \ref{HImhalodsym}, we find that satellite galaxies display typically higher gas-to-stellar mass ratios than centrals at fixed morphology, color type, and {\color{red}{group}} halo mass. It is somewhat surprising that higher satellite gas content appears to persist in rich environments{\color{red}{, except for red early types where the central and satellite values are indistinguishable. Satellites are typically expected to be quenched, but we remind the reader that  many of our HI gas masses are estimated using optical colors, and this photometric gas fraction technique may be unreliable in dense environments \citep{Cortese11}.}} Smoothed density field trends are similar to but weaker than the group halo mass trends illustrated in Fig.\ \ref{HImhalodsym}. {\color{red}{As illustrated in Fig. \ref{HImhalombaryfrac}, these environmental effects on galaxy gas fractions are not solely due to the connection between galaxy mass and group halo mass; while high and low gas fraction centrals display similar group halo masses at fixed baryonic mass, high gas fraction satellites inhabit significantly lower mass groups than lower gas fraction satellites. Thus for satellites, group halo mass is a more important driver of gas richness than galaxy mass.}}

\subsection{UV Disk Growth and Environment}

\begin{figure}
\epsscale{1.15}
\plotone{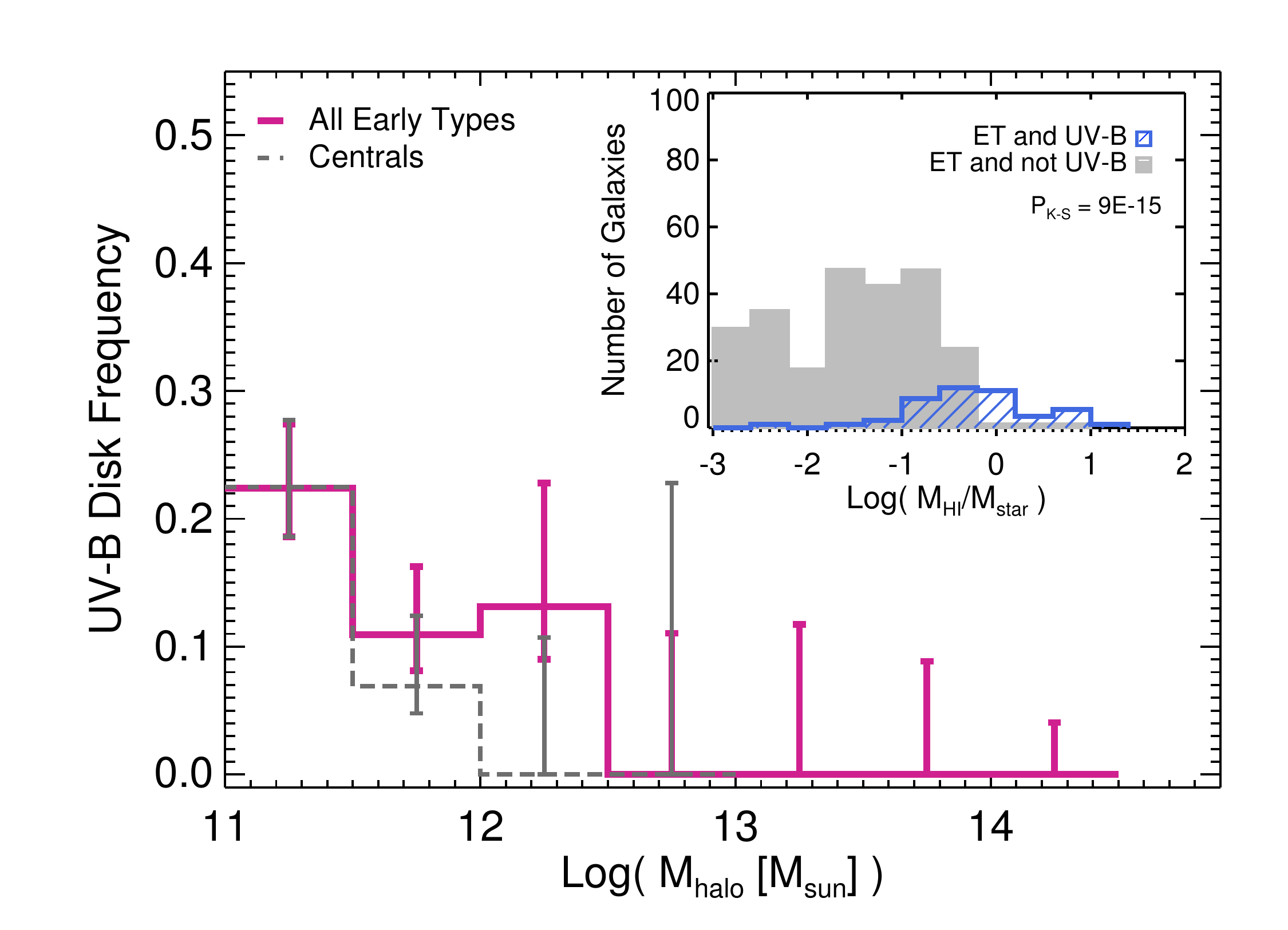}
\caption{Frequency and gas content of early-type UV-B disk hosts in the ECO+G sample. The magenta line indicates the frequency of UV-B disks in early types as a function of group halo mass, and the grey dashed line indicates UV-B disk frequency among early-type centrals alone. The inset shows the distribution of HI gas-to-stellar mass ratios for UV-B and non-UV-B early types. HI gas masses are derived as described in \S \ref{HIest}. A strong preference towards higher gas-to-stellar mass ratios is observed for early-type UV-B disk hosts.}
\label{UVBfreqs}
\label{UVBHI}
\end{figure}

  In the ECO+G subsample, we find that UV-B disks are relatively common (Fig.\ \ref{UVBfreqs}), occurring with $34^{+1.5}_{-1.4}\%$ overall frequency. Fig.\ \ref{UVBfreqs} shows that {\color{red}{the frequency of UV-B disks in early types is greatest at low group halo masses. UV-B disk frequencies among early type galaxies only reach measurable values in our sample below $M_{halo} \sim 10^{12} M_{\odot}$, approximately the same group halo mass scale below which blue-sequence early types and gas-dominated galaxies emerge. Correspondingly, we find that UV-B disks are most common in early types at low galaxy mass. The $\sim30\%$ frequency of UV-B disks we find among early types at low baryonic masses (see Fig.\ \ref{growingfreq}) is consistent with the reported $42^{+9}_{-8}\%$ UV-B disk frequencies of \citet{Me12} for early-type galaxies in a low-to-intermediate stellar mass sample. Within the mass ranges we probe, UV-B disk hosts below $M_{halo} \sim 10^{12} M_{\odot}$ are typically central galaxies,}} but additional satellite UV-B hosts with lower masses could also exist in these environments. Early type galaxies hosting UV-B disks show a strong tendency to host larger HI gas reservoirs than galaxies without UV-B disks (K-S test $P_{same} \sim 9 \times 10^{-15}$; see Fig.\ \ref{UVBHI}){\color{red}{\footnote{Note that this result also holds with similarly high significance if we restrict only to galaxies in the ECO+A subsample.}}}, in agreement with results from \citet{Me12} for low-to-intermediate mass early type galaxies. Overlapping in their typically gas-rich, low group and galaxy mass nature, blue-sequence early-type galaxies are often UV-B disk hosts, with $74^{+5}_{-7}\%$ of blue-sequence E/S0s in the full ECO+G subsample hosting UV-B disks, or $77^{+5}_{-8}\%$ in the low baryonic mass regime below $\sim10^{10}M_{\odot}$.

\section{Discussion}
\label{disc}

  In this section, we compare our results on relationships between galaxy properties and environments to previous results in the literature and to galaxy evolution scenarios.

\subsection{Morphology-Environment Relations}

\subsubsection{Comparisons to Previous Results}
  In general, we find that in the traditional P(M$|$E) morphology-environment relation formulation, our measured early and late type frequencies behave in a manner similar to that observed in previous studies of this relation, for example, with late-type galaxy frequencies decreasing from $\sim$80\% in the least rich environments to much smaller frequencies in the most rich environments (e.g., \citealp{Dressler80}; \citealp{PG84}; \citealp{Whitmore93}). Examining early-/late-type frequencies as a function of group halo mass specifically, both \citet{Bamford09} and \citet{Hoyle12} find approximately constant frequencies in the high group halo mass regime ($\gtrsim10^{13} M_{\odot}$), which we do not observe in the {\color{red}{ECO sample, even if we select only high baryonic mass galaxies to best compare with these studies (although the error bars are large as discussed further below)}}. In addition to their default estimates for group mass based on virial radius measures, \citet{Bamford09} specifically test the use of summed luminosities as a proxy for group mass, using a definition close to that we employ, and find a very slightly more pronounced trend in the early-type fraction with this proxy, although still weaker than our {\color{red}{full sample}} trend. Similarly, \citet{Poggianti09} find no significant frequency trend with cluster velocity dispersion but do find a trend with another proxy for group halo mass, X-ray luminosity. The typical galaxy masses considered by these authors are higher than those considered in ECO, and when restricting to $M_{bary} > 10^{10} M_{\odot}$ galaxies alone, our early-/late-type frequency trends at the highest {\color{red}{group}} halo masses are relatively weak given the large error bars, except in the highest {\color{red}{group halo}} mass bin (see Fig.\ \ref{MDAmass}). \citet{Calvi12} also find variations with galaxy mass, with their intermediate stellar mass galaxies showing similar morphological mixes in all environments except in the most massive clusters, which is compatible with the behavior of our $M_{bary} > 10^{10} M_{\odot}$ subsample. \citet{Bamford09} likewise report that the form of the morphology-environment relation is strongly dependent on the stellar masses of the galaxies considered, with shifts in the overall frequency levels between subsamples. We find trends similar to the Bamford et al. results, where morphology-environment relations are similar in shape but offset between low and high mass subsamples in the sense that late-type frequencies are typically higher among low-mass galaxies (see Fig.\ \ref{MDAmass}). This offset may imply that for $M_{bary} < 10^{10} M_{\odot}$ galaxies, disks are either destroyed less frequently or regenerated more frequently than for higher mass galaxies.

\subsubsection{Morphology-Environment Relations and Disk Regrowth}

  We next consider the specific question of whether or not morphology-environment relations operate in a manner consistent with the presence of disk regrowth. If morphological transformation operates primarily towards the destruction of disks in certain regimes but towards both the destruction and regrowth of disks in others, we would expect the balance of galaxy morphological types to differ in these regimes. In a scenario where large-scale gas accretion, whether arriving cold or hot, can fuel disk regrowth, the significance of such accretion is typically theorized to depend on the halo mass of the group in which a galaxy resides (e.g., \citealp{BD03}; \citealp{Keres05}; \citealp{Nelson13}). Thus, in such a scenario, the balance of galaxy morphological types might naturally be expected to shift as a function of group halo mass. As previously mentioned, in the traditional morphology-environment relation, P(M$|$E), we observe a changing balance between early and late types as a function of group halo mass, in the sense that late types become more prevalent with decreasing halo mass. This trend is the sense in which the early/late type balance would be expected to vary if disk regrowth were to preferentially occur at low group halo mass, however it is also certainly the sense in which one would expect the relation to vary if disks are typically destroyed/quenched at high but not low group halo mass. Thus, while consistent with a halo-mass dependent disk regrowth scenario, the observed traditional morphology-environment relation does not clearly constrain its existence. We also note that the balance between early- and late-type frequencies we observe in the ECO sample varies relatively smoothly as a function of group halo mass, as has often been observed by other authors. This smooth variation could indicate a lack of sharp transitions in the onset of morphological transformation processes at particular mass scales, but alternatively it could imply that the traditional relation, in lumping all early and all late types together, washes out possible sharper trends that may occur for subpopulations of galaxies.

Considering this question from the perspective of an alternative formulation of the morphology-environment relation, P(E$|$M), another expectation of the disk regrowth model emerges. If disk regrowth were to proceed from blue early to blue late types in a particular regime, then the typical environments of blue early and blue late types in that regime should be similar as these galaxies would represent snapshots of pre- and post-transition states. Such behavior was hinted at in the observation of a possible ``inverse morphology-density relation'' at low stellar masses by KGB. As illustrated in Figs.\ \ref{mhalombarydsym} and \ref{densmstar}, we find that blue early and blue late type galaxies in the low baryonic mass regime below $\sim10^{10}M_{\odot}$ inhabit environments with similar typical group halo masses at constant baryonic mass and with typical environmental \emph{densities} of blue early types similar or lower than those of blue late types. The P(E$|$M) formulation of the morphology-environment relation then appears to be consistent with the scenario of disk regrowth in the low baryonic mass regime. If gas accretion adds to both galaxy baryonic and overall {\color{red}{group}} halo mass during the regrowth process, individual galaxies could move along the $M_{halo}$ vs.\ $M_{bary}$ relation, but relatively significant changes in typical population properties would be necessary to make the blue early and blue late type populations distinct in this space given the scatter within each population. The trend towards lower number density environments for blue early types compared to blue late types possibly points to a post-merger status for blue early types. In the spirit of the alternative formulation of the morphology-environment relation, consideration of the typical environments of various subclasses of galaxies leads to further insights regarding environmental thresholds in disk growth as discussed in the next section.

\subsection{The Regime of Extreme Gas Richness and Recent Disk Growth}

\begin{figure*}
\epsscale{1.15}
\plotone{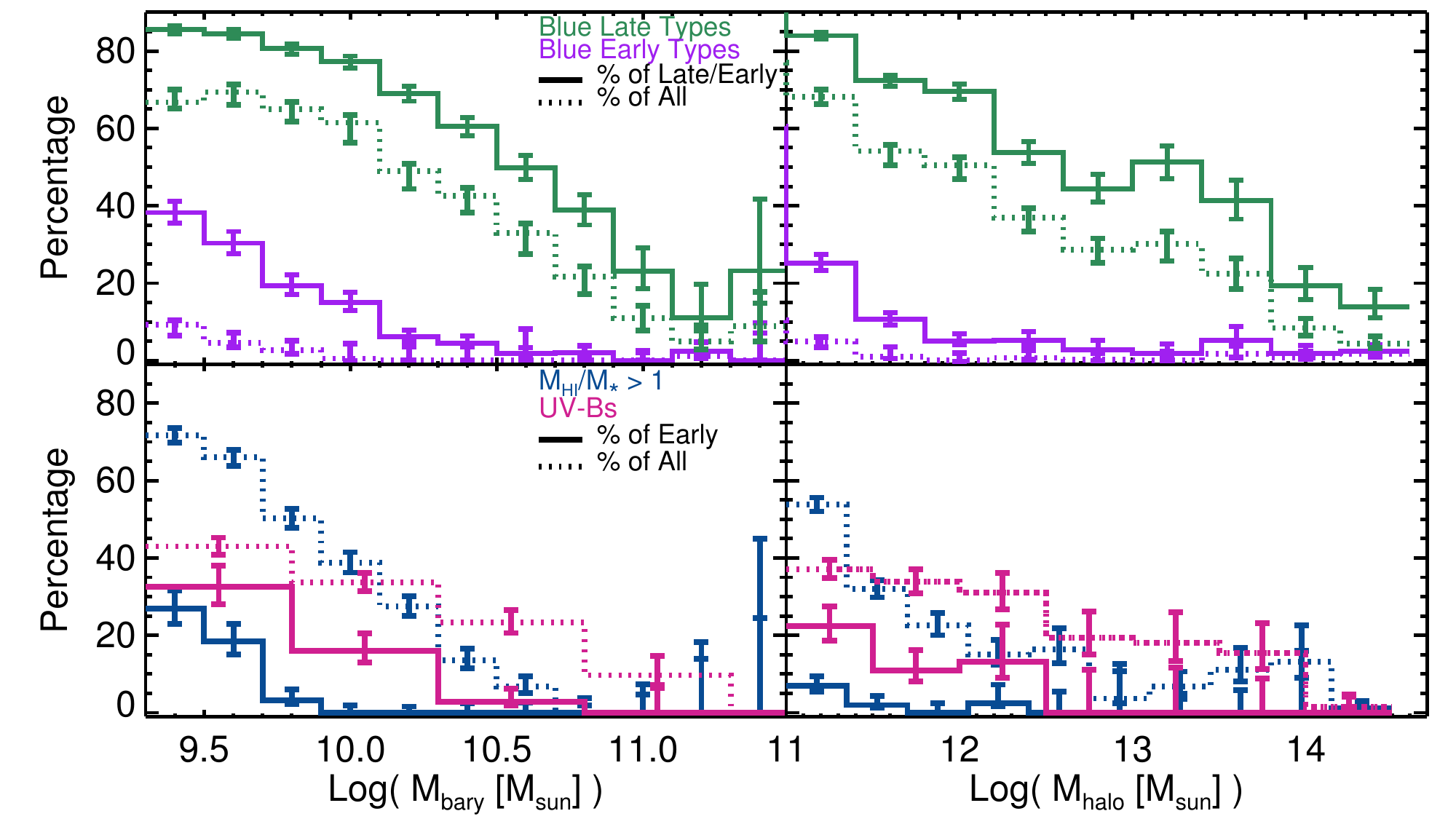}
\caption{Illustration of the frequency (percentage) of various galaxy populations with the potential for disk growth as a function of galaxy and group halo mass. The top panels indicate the frequency of blue early- and late-type galaxies as a fraction of all galaxies with the same morphology and of all types. The lower panels show the frequency of gas-dominated and UV-B disk galaxies among all galaxy types and early types specifically. While each of these populations clearly displays a unique variation in frequency with both galaxy mass and environment, each of these ``growing'' populations occurs most commonly where galaxy masses and group halo masses are low.}
\label{growingfreq}
\end{figure*}

We have presented multiple results that add up to the impression that extremely gas rich galaxies and those potentially regrowing disks are preferentially found in the low group halo mass ($M_{halo} \lesssim 10^{11.5} M_{\odot}$) and low galaxy mass ($M_{bary}$ or $M_{*} \lesssim 10^{10} M_{\odot}$, near the ``gas-richness threshold'' stellar mass at $M_{*} \sim 10^{9.7} M_{\odot}$) regime. We note that this regime is approximately defined, as in some cases we observe a continuum of galaxy properties in our sample, whereas in others, typically involving central galaxies, we see more abrupt transitions in properties between well-defined regimes. {\color{red}{In Fig.\ \ref{growingfreq}, we summarize the galaxy mass and group halo mass distributions of the relevant galaxy populations. In ECO, blue-sequence early types and early types with $M_{HI}/M_{*} > 1$ occur most commonly below a group halo mass of $\sim10^{11.5} M_{\odot}$. This group halo mass regime is also where extreme gas-to-stellar mass ratios commonly emerge in our sample, with a relatively sharp uptick as a function of decreasing group halo mass as can be seen in the lower right panel of Fig.\ \ref{growingfreq}. Likewise, early-type UV-B galaxies are most common in the same low group halo mass and low galaxy mass regimes, albeit with a broader distribution extending to somewhat higher masses as well. Early types with UV-B disks are more often gas rich than early types without UV-B disks, and most low-mass blue-sequence early types in the ECO+G subsample host UV-B disks.}}

A possible explanation for this constellation of results is that this low galaxy mass and low group halo mass region of parameter space represents a preferred regime where gas is abundantly available to galaxies, fueling star formation that allows many early types to live on the blue sequence, develop UV-B disks, and potentially even regrow larger disk structures. The existence of such a gas and star formation rich regime could be a symptom of a large-scale cosmological accretion process that is particularly efficient at supplying gas into galaxies at low mass scales. One such theorized process is ``cold-mode'' gas accretion, thought to preferentially act at group halo mass scales below this $\sim10^{11.5} M_{\odot}$ mass scale at $z\sim$0 (e.g., the $\sim10^{11.3} - 10^{11.5} M_{\odot}$ scale of \citealp{Keres05}; \citealp{Keres09}). This model has recently been challenged by simulation results using the AREPO code, which imply that cold-mode accretion is not as significant for galaxies residing in low mass halos as previously thought, however these results suggest that accretion of heated gas is a more significant contributor in this regime, causing the level of gas accretion into low mass halos to remain high \citep{Nelson13}. As seen in the simulations of \citet{Zehavi12}, from $z\sim$1 to the present the bulk of the stellar mass growth in low-mass halos ($M_{halo} \lesssim 10^{12} M_{\odot}$) is still due to star formation, while in high-mass halos, such growth is mainly due to mergers. 

 Within the galaxy baryonic mass range we consider, we find that blue early-type and UV-B disk host galaxies are typically centrals in the $M_{halo} \lesssim 10^{11.5} M_{\odot}$ regime (see Figs.\ \ref{MDB}b and \ref{UVBfreqs}). We also find that the strong uptick in gas-dominated galaxy fraction in this regime is primarily a central galaxy phenomenon, which may point towards accretion fueling of central galaxies in these environments (see Fig.\ \ref{HIfraccs}). However, we also note that in the lowest {\color{red}{group}} halo mass environments we probe there are relatively few satellite galaxies within our baryonic mass range, and therefore it is plausible that gas-dominated, disk-growing satellites with lower masses could be common in such environments as well. From simulations, typical $z\sim$0 gas accretion rates for satellite galaxies may be lower than for central galaxies (e.g., \citealp{Keres09}), but gas accretion may still play an evolutionarily significant role for satellites (e.g., \citealp{DB06}; \citealp{Simha09}).

\section{Conclusions}
\label{conc}
  In this work, we have considered two primary galaxy samples, the Environmental COntext (ECO) catalog, and the B-semester region of the REsolved Spectroscopy Of a Local VolumE (RESOLVE) survey. Both samples reach into the high-mass dwarf galaxy regime and span a variety of environments, with the larger ECO catalog sample including the greatest environmental diversity. Through comparison to the more complete RESOLVE-B catalog, we apply corrections for incompleteness effects in ECO, creating an approximately baryonic mass limited catalog down to $10^{9.3} M_{\odot}$. In this analysis, we have employed high-quality, custom-reprocessed optical, near-IR, and UV photometry along with morphological classifications, atomic gas mass estimates, and multiple metrics of galaxy environment.

Our key results are as follows.

\begin{itemize}

\item We observe a traditional morphology-environment relation, P(M$|$E), similar to the expected form but with offset amplitudes between the low and high baryonic mass galaxy samples in the sense that late types are more common at low mass.

\item We find the form of the traditional morphology-environment relation to be consistent with the scenario that morphological transformation from early to late types (disk regrowth) could occur in a preferred low group halo mass regime. {\color{red}{However, this relation does not strongly constrain the existence of this scenario, as the relation can also be explained by the occurrence of disk destruction/quenching at high but not low group halo mass.}}

\item We consider an alternative form of the morphology-environment relation, P(E$|$M), which is instructive as a way of quantifying the \emph{typical} environments of galaxies of various classes. This formulation leads to the observation that typical blue early-type and blue late-type galaxy group halo masses are similar at constant baryonic mass, which is again consistent with expectations from the disk regrowth scenario. Likewise, the typical environmental densities of blue early types are similar or lower than those of blue late types at constant baryonic mass, potentially reflecting a post-merger state for blue early types.

{\color{red}{
\item The P(E$|$M) formulation of the morphology-environment relation also reveals that for central galaxies, there is no discernible relationship between group halo mass and morphology at fixed galaxy mass: the typical halo masses for all early-type and all late-type centrals of a given mass are the same. However, satellite galaxies in different color classes divide strongly in their typical group halo masses at fixed galaxy mass, where red early and late types occupy higher group halo mass environments than both blue early and late types. These observations suggest the traditional morphology-environment relation is largely driven by a relationship between morphology and galaxy mass for centrals and by a relationship between color and environment for satellites.
}}
\item We find that the low group halo mass regime below $\sim10^{11.5} M_{\odot}$ is associated with the emergence of blue-sequence early types, gas-dominated galaxies, and early-type UV-Bright disk hosts as common contributors to galaxy populations. These three sub-populations are closely linked in this regime, implying the low group halo mass regime is a preferred regime for ongoing, significant disk growth.

\end{itemize}

These results lend strong support to the idea that theorized morphological transformation from early to late types can occur, particularly where galaxy and group halo masses are low. {\color{red}{Thus, even if galaxy disks are destroyed through mergers and interactions in this regime, they may have significant opportunity to regrow, whereas galaxy disk regrowth does not appear likely in higher group halo mass environments}}. To investigate even more direct signatures of disk regrowth, we next turn to the examination of detailed early-type galaxy kinematics in the context of the RESOLVE survey, where the availability of such kinematic information combined with the type of environmental information considered here creates a unique opportunity for understanding the connection between galaxy properties on small and large scales.

\acknowledgments

We would like to thank R. Gonzalez, N. Padilla, S. Khochfar, {\color{red}{J. Cisewski, A. Baker, and E. Feigelson for helpful discussions. We also thank the anonymous referee for helpful comments that have led to significant improvements in the presentation of this work.}} AJM acknowledges funding support from a NASA Harriett G. Jenkins Fellowship, a University of North Carolina Royster Society of Fellows Dissertation Completion Fellowship, a North Carolina Space Grant, and \emph{GALEX} GI grants NNX07AT33G and NNX09AF69G. SJK, KDE, DVS, and MAN acknowledge support from the NSF CAREER grant AST-0955368. KDE and DVS acknowledge additional support from GAANN Fellowships and North Carolina Space Grants.

Funding for SDSS-III has been provided by the Alfred P. Sloan Foundation, the Participating Institutions, the National Science Foundation, and the U.S. Department of Energy Office of Science. The SDSS-III web site is \verb1http://www.sdss3.org1. SDSS-III is managed by the Astrophysical Research Consortium for the Participating Institutions of the SDSS-III Collaboration including the University of Arizona, the Brazilian Participation Group, Brookhaven National Laboratory, University of Cambridge, Carnegie Mellon University, University of Florida, the French Participation Group, the German Participation Group, Harvard University, the Instituto de Astrofisica de Canarias, the Michigan State/Notre Dame/JINA Participation Group, Johns Hopkins University, Lawrence Berkeley National Laboratory, Max Planck Institute for Astrophysics, Max Planck Institute for Extraterrestrial Physics, New Mexico State University, New York University, Ohio State University, Pennsylvania State University, University of Portsmouth, Princeton University, the Spanish Participation Group, University of Tokyo, University of Utah, Vanderbilt University, University of Virginia, University of Washington, and Yale University. 

Based on observations made with the NASA Galaxy Evolution Explorer.
\emph{GALEX} is operated for NASA by the California Institute of Technology under NASA contract NAS5-98034.

This publication makes use of data products from the Two Micron All Sky Survey, which is a joint project of the University of Massachusetts and the Infrared Processing and Analysis Center/California Institute of Technology, funded by the National Aeronautics and Space Administration and the National Science Foundation.



{\it Facilities:} \facility{Sloan}, \facility{GALEX}, \facility{2MASS}.

\bibliographystyle{apj}
\bibliography{thesis}

\clearpage
\begin{sidewaystable}
\centering
\tiny
\caption{ECO Sample Properties}
\label{tbl1}
\begin{tabular}{cccccccccccccccccccccc}
\tableline\tableline 
\hline
\\[0.25pt]

Gal. ID & RA  & Dec & cz & $M_{r}$ & $\log${$M_{*}$/$M_{\odot}$} & $(u-r)^{e}$ & $(u-J)^{m}$ & $R_{50\%}$ & $R_{90\%}$ & M & $F_{M}$ & Grp. ID & Grp. cz & $F_{C}$ & $\log${$M_{halo}$/$M_{\odot}$} & Dens. & $F_{A}$ & $F_{HI}$ & $F_{G}$ & $CC_{r}$ & $CC_{b}$\\
& (deg) & (deg) & (km/s) & & & & & ($''$) & ($''$) & & & &(km/s)& & & & & & & &\\

\noalign{\smallskip}\hline\noalign{\smallskip}
ECO12079&130.065&  23.375&7230.9&-18.80& 8.91& 0.88& 2.06& 3.5& 8.9&L &1&8480&7230.9&1&11.02& 0.19&0&1&0&1.14&1.00\\
ECO02120&130.095&  23.539&3534.8&-20.82&10.23& 1.31& 3.31&16.2&34.2&L &2&1918&3534.8&1&11.63& 0.24&0&1&0&1.07&1.00\\
ECO08137&130.106&  18.316&4373.9&-17.87& 8.95& 1.84& 3.42& 4.1&13.6&E &2&5958&4373.9&1&10.84& 0.67&0&1&0&1.15&1.00\\
ECO00569&130.113&  18.574&4491.7&-18.03& 8.81& 0.76& 1.97&12.0&30.7&L &1& 552&4491.7&1&10.87& 0.74&0&1&0&1.15&1.00\\
ECO10139&130.126&  27.631&6151.0&-18.99& 9.23& 1.18& 2.30&13.4&27.7&L &2&7240&6151.0&1&11.06& 0.57&1&2&0&1.10&1.00\\
ECO09648&130.153&  41.901&7411.4&-18.11& 8.79& 1.10&18.20& 5.1&11.8&L &2&6898&7411.4&1&10.88& 0.44&0&1&0&1.16&1.00\\
ECO05187&130.161&  27.241&5488.7&-19.49& 9.91& 2.01& 3.87& 9.8&28.7&L &2&3962&5488.7&1&11.18& 0.39&1&1&0&1.04&1.00\\
ECO10274&130.183&  37.010&6827.2&-18.66& 9.01& 0.80& 3.38& 9.8&18.7&L &1&7336&6827.2&1&10.99& 0.27&0&1&0&1.07&1.00\\
ECO00776&130.188&  30.479&5596.4&-18.16& 8.51& 0.73& 1.92& 4.2& 9.2&L &1& 742&5596.4&1&10.90& 0.28&0&1&0&1.17&1.00\\
ECO00777&130.190&  13.707&4369.5&-17.51& 8.53& 0.99& 1.96& 7.0&16.6&L &1& 743&4369.5&1&10.77& 0.16&1&2&0&1.51&1.00\\
\tableline
\end{tabular}
\tablecomments{Table \ref{tbl1} contains all ECO galaxies with $M_{r} < -17.33$. A portion of Table \ref{tbl1} is shown here for guidance regarding its form and content; this complete table will be made available as a machine-readable file (see http://resolve.astro.unc.edu/pages/data.php). Velocities (cz) are given in local group corrected form. M indicates early or late type morphology, and $F_{M}$ is a flag that indicates the source of the classification: 1 for quantitative cut and 2 for by eye from \citet{Galzooclass}. Group ID numbers, center velocities (local group corrected), central/satellite designations, and {\color{red}{group}} halo masses are given by the Grp. ID, Grp. cz, $F_{C}$ (1 indicates central), and $\log${$M_{halo}$/$M_{\odot}$} columns. Dens. indicates normalized environmental density smoothed $\sim$1.43 Mpc scales. $F_{A}$ is a flag with value equal to 1 where a galaxy belongs to the EC0+A sample. $F_{HI}$ is a flag indicating whether: (1) photometric gas estimates or (2) gas measurements were used for each galaxy. $F_{G}$ is a flag with value equal to 1 where a galaxy belongs to the EC0+G sample. $CC_{r}$ indicates the sample membership completeness correction factor that accounts for redshift catalog incompleteness, and $CC_{b}$ indicates the additional completeness correction factor that accounts for incompleteness due to boundary effects.}

\end{sidewaystable}

\end{document}